\documentclass[aip,jcp,a4paper]{revtex4-1}
\usepackage{amssymb}
\usepackage{graphicx}
\usepackage[normalem]{ulem}
\usepackage[usenames]{color}

\begin{document}

\title{Quantum Monte Carlo estimation of complex-time correlations for
the study of the ground-state dynamic structure function}
\author{R. Rota}
\affiliation{Dipartimento di Fisica and INO-CNR BEC Center, Universit\`a degli Studi di Trento, 
I-38123 Povo, Trento, Italy}

\author{J. Casulleras}
\author{F. Mazzanti}
\author{J. Boronat}
\affiliation{Departament de F\'{i}sica i Enginyeria Nuclear, Universitat Polit\`{e}cnica 
de Catalunya, Campus Nord B4-B5, E-08034, Barcelona, Spain} 

\begin{abstract}

We present a method based on the Path Integral Monte Carlo formalism for
the calculation of ground-state time correlation functions in quantum
systems. The key point of the method is the consideration of time
as a complex variable whose phase $\delta$ acts as an adjustable parameter.
By using high-order approximations for the quantum propagator, it
is possible to obtain Monte Carlo data all the way from purely imaginary
time to $\delta$ values near the limit of real time. As a consequence,
it is possible to infer accurately the spectral functions using simple
inversion algorithms. We test this approach in the calculation
of the dynamic structure function $S(q,\omega)$ of two one-dimensional
model systems, harmonic and quartic oscillators, for which  $S(q,\omega)$
can be exactly calculated. We notice a clear improvement in the calculation
of the dynamic response with respect to the common approach based
on the inverse Laplace transform of the imaginary-time correlation
function.



\end{abstract}

\maketitle

\section{Introduction}

In the last decades, quantum Monte Carlo (QMC) methods have been
extensively used in the field of quantum many-body  physics. Many of these
numerical techniques rely on stochastic propagation in imaginary  time and 
can provide extremely accurate results for the thermodynamic and static
properties of many-body  systems, even in those where quantum correlations
make  unavoidable the use of non-perturbative
approaches.\cite{CeperleyRMP,Sarsa00,electrongas,unitary} The main drawback
of QMC methods is the difficulty arising in the calculation of spectral
functions. These functions, which are  particularly relevant for the study
of the dynamical properties of quantum many-body systems (e.g. the
excitation  spectrum or the transport coefficients), can be obtained as 
Fourier transforms of real-time correlation functions. A QMC calculation of
these  quantities, however, is particularly inefficient since the rapidly
oscillating exponentials appearing in the definition of real-time
propagators make the statistical errors grow exponentially with time. Many
approximation schemes have been developed and used to  investigate the
dynamic properties of quantum many-body systems. For instance,
centroid~\cite{cao} or ring-polymer molecular dynamics\cite{craig} has been
successfully applied to the study of quantum many-body systems in the
semi-classical regime.  In the limit of zero temperature, an alternative
approach is to use correlated perturbation theory\cite{krotscheck} relying
on the ground-state properties of the system obtained with QMC
calculations.~\cite{mazzanti,Macia12,krotscheck2}

Nevertheless, the mainstream approaches to the calculation of spectral
functions from QMC simulations consist in attempting a  numerical inversion
of a Laplace transform. This integral transform relates the
desired spectral  functions to the correlation functions in imaginary
time, easily attainable with QMC methods. However, the inverse Laplace
transform of noisy data is an ill-posed problem. This means that, given a
particular set of data for the imaginary-time correlation function, it is
hardly possible to recover a unique, well-defined solution to the problem. 
Sophisticated regularization techniques can then be used to reproduce a
reasonable estimate of the spectral function.~\cite{InvProbTextbook} In the last decades, several algorithms to deal with the inverse Laplace
transform of noisy data have been proposed,~\cite{Jarrell96,Boninsegni96,Sandwik98,Mishchenko00,Vitali10} but these methods can only be reliably applied to the analysis of the low-energy dynamic properties of quantum systems,
since the Laplace kernel tends to suppress high-energy contributions. In order to overcome these limitations and to get more accurate results of spectral functions from
QMC data, it is necessary to develop new estimators for the quantum time
correlation functions. ~\cite{Roggero13}

In this work, we propose to infer the dynamic structure function of a
quantum system at zero temperature from a QMC estimation of the
corresponding correlation
function in complex time. Similar approaches have been already used for
studying the dynamic properties of quantum systems at finite temperature
$T$. In this case, the $e^{-\beta\hat{H}}$ term (with $\beta =
1/T$) appearing in the definition of the thermal averages can be considered
as an evolution operator in imaginary time. Thus, the real-time correlation
function can be rewritten in terms of a correlation function in complex
time,~\cite{Schofield,Thirumalai84} which can be calculated using
path-integral formalism\cite{Feynman} and estimated in QMC
calculations.\cite{Chakrabarti98} Even though this estimation is reliable
only for times $t \lesssim \hbar \beta$, the spectral functions obtained
within this approach exhibit a significant improvement over the results
derived from analytic continuation of imaginary-time correlation
functions.~\cite{Krilov01,Sim01,Krilov01CP,Nakayama06,Kegerreis08,Jadhao08,Bonella10a,Bonella10b} 

Our goal is to extend this formalism to the calculation of ground-state
time correlation  functions, even considering that at zero temperature the
notion of complex time has not a precise physical meaning. 
This strategy allows us to introduce an adjustable
parameter, namely the phase $\delta$ of the complex time $t_c =
|t_c|e^{-i\delta}$, which makes possible to calculate the correlation
function in an intermediate regime between the commonly used imaginary time
($\delta = \pi/2$) and real time ($\delta = 0$).

More precisely, we sample paths connecting two configurations distributed
according to the ground-state wave function of the quantum system and
calculate, over these paths, the propagator at the complex time $t_c$.
Changing the phase $\delta$, we can find an optimal value for which the
correlation functions estimated with QMC are affected by moderate
statistical errors and, at the same time,  present a relevant amount of
information on the real dynamics of the quantum system. This approach makes
it possible to infer the spectral functions using rather simple 
inversion techniques since the ill-posed character of the inversion
procedure is appreciably reduced. In this way,  more accurate and more
stable results than the usual ones, based on the inverse Laplace transform
of imaginary-time data, can be obtained.

Similarly to what happens in the
case at finite temperature, the QMC estimation of the ground state
correlation function in complex time is reliable only up to a certain value
of $|t_c|$ depending on $\delta$, above which the statistical error becomes
too large and makes the numerical results meaningless. It is therefore
crucial to develop strategies that make the range of accessible times as
large as possible. In this work, we propose to tackle this problem using
high-order approximations for the quantum propagator.~\cite{chin} In
particular, we show 
that the propagator derived by Zillich \textit{et al.}~\cite{Zillich10} 
is particularly suitable for the complex-time evolution.

The rest of the paper is organized as follows. In Sec. II, we discuss the
QMC method that we have devised for the calculation of complex-time correlation
functions and, more briefly, the inversion method that we have used to
obtain the dynamic structure function. In Sec. III, the results
obtained with this method in one-dimensional problems are shown and
compared with the standard approach relying only on imaginary-time
correlation functions. Finally, the summary and main conclusions are
reported in Sec. IV.

\section{Method}

\subsection{Calculation of the complex-time correlation function}

The main objective of our work is the calculation of ground-state time
correlation functions of a quantum system. 
At zero temperature, a general time correlation function is defined as 
\begin{equation}\label{eq:DefinitionC}
C_{AB}(t_c) = \langle \Psi_0 | e^{it_c \hat{H} / \hbar} \hat{A} e^{-it_c \hat{H} / \hbar} \hat{B} | \Psi_0 \rangle \ ,
\end{equation}
where $\hat{A}$ and $\hat{B}$ are time-independent quantum mechanical operators
in the Schr\"odinger picture corresponding to measurable observables, 
$\hat{H}$ is the Hamiltonian, and
$|\Psi_0\rangle$ is the ground state. For the sake of
simplicity, in the following  we consider correlations among operators which
are diagonal in  coordinate  space and use one-dimensional
notation (the generalization to multi-dimensional space is straightforward).

The main idea of this work is to calculate $C_{AB}(t_c)$, defined
in Eq. \ref{eq:DefinitionC}, where $t_c$ has been analytically extended to
the complex plane. We indicate with $t_m > 0$ and $-\delta$
the modulus and the phase of the complex time, $t_c = t_m e^{-i
\delta}$, respectively. In order to elaborate a form for the estimator of $C_{AB}(t_c)$
implementable in computer simulations, we rewrite Eq. \ref{eq:DefinitionC}
in the coordinate space,
\begin{eqnarray}
C_{AB}(t_c) & = & \int dx_0 dx_M \, e^{it_c E_0} \langle \Psi_0 | x_M
\rangle \langle x_M | \hat{A} e^{-it_c \hat{H}} \hat{B} | x_0 \rangle \langle x_0 | \Psi_0 \rangle \ = \nonumber \\
& = & \mathcal{N} \int dx_0 dx_M \, \Psi_0^\star(x_M) A(x_M) G(x_0,x_M;t_c) B(x_0) \Psi_0(x_0) \ , \label{eq:DefinitionCwithGSWF}
\end{eqnarray}
where  $G(x_0,x_M;t_c) =
\langle x_M | e^{-it_c \hat{H}} | x_0 \rangle$ is the propagator from 
position $x_0$ to position $x_M$ in complex time $t_c$, and
$\mathcal{N}$ is a normalization constant.  In
the general case of complex time $t_c$, the propagator $G(x_0,x_M;t_c)$
is a complex function that becomes  real and positive  
only when $t_c$ is a purely imaginary time. Thus, the function
$G(x_0,x_M;t_c)$ cannot be used as a probability distribution function
for the sampling of coordinates in any QMC algorithm (as it is normally
done, for instance, in the PIMC method). Therefore, what we do is to sample
first the positions $x_0$ and $x_M$ according to a probability
distribution constructed from an accurate approximation to the ground-state
wave function. This sampling can be performed using any conventional QMC
technique at zero temperature. In this work, we use the Path Integral Ground
State (PIGS) method.~\cite{Sarsa00,Rota10} Then, having sampled the positions $x_0$ and
$x_M$, we calculate $C_{AB}(t_c)$ estimating the quantity $A(x_M)
G(x_0,x_M;t_c) B(x_0)$. 

In order to carry on this procedure one 
needs to know the exact form of the Green's function $G(x_0,x_M;t_c)$ 
for any value
$t_c$, but this is in general unknown. However, what is possible is to construct 
accurate approximations to the propagator
in the limit of small $t_m = |t_c|$. Then, to estimate
$C_{AB}(t_c)$ for larger values of $t_m$ we use the path-integral
formalism to rewrite $G(x_0,x_M; t_c)$ as a convolution of
$M$ propagators of a shorter time $\varepsilon_c = t_c/M$,
\begin{equation}\label{eq:Convolution}
G(x_0,x_M;t_c) = \int dx_1 \ldots dx_{M-1} \prod_{k=1}^M
G\left(x_{k},x_{k-1};\varepsilon_c \right) \ .
\end{equation}

Within this approach, it becomes necessary to sample all the configurations
$\{ x_1, x_2, \ldots ,x_{M-1} \}$, i.e., to build paths from the
position $x_0$ to the position $x_M$. However, the choice of the probability distribution $p_{\text{path}}(x_0,x_1,\ldots,x_M)$ for these paths is not trivial and depends on the system studied. Generally, we notice that using imaginary-time propagator to this end is not a good choice, because in this case the sampled paths would remain close to the minimum energy path and the estimator would not be able to capture all the
contributions to $C_{AB}$ coming from the excited states. 
As a  simple and flexible enough option, it is possible to choose $p_{\text{path}}$ as 
the product of $M$ free propagators of imaginary-time step $\tau_s$,
\begin{equation}\label{eq:ppath}
p_{\text{path}}(x_0, x_1,\ldots,x_M) = \prod_{k=1}^M
G_{\text{free}}(x_k,x_{k-1};\tau_s) \ ,
\end{equation}
with 
\begin{equation}\label{eq:free}
G_{\text{free}}(x_k,x_{k-1};\tau_s) = (4 \pi \lambda \tau_s)^{Nd/2}
\exp\left( -\frac{(x_k-x_{k-1})^2}{4\lambda\tau_s} \right)  \ .
\end{equation}
In Eq. \ref{eq:free}, $N$ is the number of particles, $d$ is the
dimensionality of the system, and $\lambda = \hbar^2/(2m)$. This choice
indeed allows to construct the paths by means of simple sampling 
techniques which do not require a large computational effort, like for
instance the staging algorithm.~\cite{stag1,stag2} In the case of
quantum systems interacting with a smooth potential, we notice that it is
possible to obtain good results for $C_{AB}(t_c)$ using $p_{\text{path}}$
in Eq. \ref{eq:ppath}, provided that the parameter $\tau_s$ is properly
chosen.  Indeed, we see that the variance of the estimator for $C_{AB}(t_c)$
is reduced when the free propagator in the imaginary time $\tau_s$ is
similar to the modulus of the kinetic propagator in the complex time
$\varepsilon_c$.

Since the purpose of this work is to test our QMC approach  in  two model
systems interacting with smooth potentials (the quantum harmonic and
quartic oscillators), we decide to use this choice of $p_{\text{path}}$ with
$\tau_s \simeq (\Re [1/(i \varepsilon_c)])^{-1}$ to perform the sampling of
the paths $\{ x_1, x_2, \ldots ,x_{M-1} \}$. Nevertheless, 
this may not be the best choice in general, and one may have
to use more sophisticated and more
computationally demanding algorithms for the sampling of the paths.

Once the probability distribution $p_{\text{path}}$ is chosen, 
the expression of the ground state complex time correlation function becomes
\begin{eqnarray}
C_{AB}(t_c) & = & \mathcal{N'} \int dx_0 \ldots dx_M A(x_M) \frac{\prod_{k=1}^M G(x_{k},x_{k-1};\varepsilon_c)}
{p_{\text{path}}(x_0, x_1,\ldots,x_M)} B(x_0) \times \nonumber \\
& & \Psi_0(x_M) p_{\text{path}}(x_0, x_1,\ldots,x_M) \Psi_0(x_0) \ . \label{eq:DefinitionCImportanceSampling}
\end{eqnarray}

At this point, one has to choose an approximation scheme for
$G(x_k,x_{k-1};\varepsilon_c)$ in order to derive an analytical expression
that can be implemented in computer simulations. Increasing the number of
convolution terms $M$, and thus decreasing the modulus of $\varepsilon_c$,
it is possible to systematically improve the quality of the approximation
and to asymptotically recover the exact correlation function.
Nevertheless, every propagator $G(x_{k},x_{k-1};\varepsilon_c)$
introduces an oscillating phase term in the integrand of Eq.
\ref{eq:DefinitionCImportanceSampling}, and thus the statistical noise of the
estimator for $C_{AB}(t_c)$ increases notably when $M$ becomes large. In
order to obtain reliable results,  
it is fundamental to develop numerical strategies that keep the
number $M$ of convolution terms as low as possible.

The simplest approximation to the
propagator is the  primitive approximation (PA), which relies on the
factorization $e^{i t_c \hat{H}} \simeq e^{i t_c \hat{K}} e^{i t_c
\hat{V}}$, where $\hat{K}$ and $\hat{V}$ are the kinetic and
potential operators, respectively. In this scheme, the complex-time propagator can be written as
\begin{eqnarray}
G (x_k,x_{k-1}; \varepsilon_c) & \simeq & G_{PA} (x_k,x_{k-1}; \varepsilon_c) = \nonumber \\
& = & \exp \left( - \frac{(x_k-x_{k-1})^2}{4 \lambda \, i \varepsilon_c} \right) 
\exp \left( -i \frac{V(x_k) + V(x_{k-1})}{2 \hbar} \varepsilon_c \right) \ . 
\label{eq:propagatorPrimitive}
\end{eqnarray}

The PA approximation is easily implementable within our QMC procedure but
requires a large number $M$  of convolution terms in Eq.
\ref{eq:DefinitionCImportanceSampling}. In
order to improve the accuracy, it is important to use
higher-order approximations to the complex-time propagator. In
conventional PIMC simulations, a significant improvement in efficiency  
can be obtained using symplectic expansions of 
the time-evolution operator that incorporates double commutators between kinetic and potential
operators.~\cite{TakahashiImada,chin}
For local potentials, these commutators lead to extra terms that are
exponentials of the gradient of the potential squared times the third
power of the time step.
The inclusion of this contribution in the propagator largely improves the
efficiency of the PIMC~\cite{Sakkos09} and PIGS~\cite{Rota10} methods. In
imaginary-time propagation, the contribution of the double commutator
always appears in the argument of the exponential with a negative sign.
However, in complex time this sign turns out to be positive
for $\delta<60^0$, producing largely increasing amplitudes and thus 
unreliable results that make the
use of this high-order scheme unpractical (see Appendix A). 
 Therefore, it is very important to look for other
expansions which can improve the PA but that do not include 
double-commutator terms.

A high-order approximation for the complex-time propagator 
without double commutator has been reported in Ref.
\onlinecite{Zillich10}. In that work, the authors were able to improve the
quality of the small-time propagator by introducing a linear
combination, with some negative coefficients, of different symplectic expansions
on the same time. This expansion has some drawbacks when
used in conventional PIMC simulations, since it gives rise to an
approximation for the imaginary-time propagator which is not 
positive definite. This feature does not represent a problem here, 
since in the calculation of $C_{AB}(t_c)$ the complex-time
propagator is not used as the probability distribution
of the Monte Carlo sampling but rather as the estimator.

Once we have chosen the approximation for the complex-time propagator, the
only thing that is still lacking in order to calculate $C_{AB}(t_c)$ is the normalization
constant $\mathcal{N'}$. This can be computed imposing the
autocorrelation function of the identity operator to be  $1$ for any value of
$t_c$. Therefore, if we define the complex quantity
\begin{equation}
O_A(x_0, \ldots, x_M) = \frac{ \prod_{k=1}^M G_{A}(x_k,x_{k-1};\varepsilon_c)}{p_{\text{path}}(x_0, x_1,\ldots,x_M)} \ , 
\label{eq:EstimatorPropagator}
\end{equation}
where $G_{A}(x_k,x_{k-1};\varepsilon_c)$ is the chosen approximation for
the time propagator, the complex-time correlation function in 
Eq. \ref{eq:DefinitionCImportanceSampling} can be written as 
\begin{equation}
C_{AB}(t_c) = \frac{\langle A(x_M) O_A(x_0, \ldots, x_M) B(x_0)\rangle}{\langle O_A(x_0, \ldots, x_M) 
\rangle} \ . \label{eq:CMontecarlo}
\end{equation}
The bracket $\langle \ldots \rangle$  indicates the averages over the
configurations $\{ x_0, x_1, \ldots ,x_M \}$ sampled following the
scheme described above, i.e., with $x_0$ and $x_M$ sampled according to
a reasonable approximation of the ground-state wave function, 
and $\{ x_1, x_2, \ldots ,x_{M-1}\}$ sampled 
according to the probability distribution $p_{\text{path}}(x_0, x_1,\ldots,x_M)$. 

Summarizing, the evaluation of $C_{AB}(t_c)$ (\ref{eq:CMontecarlo}) for a
given complex time $t_c = t_m e^{-i \delta}$ consists of the following steps:
\begin{enumerate}
\item{To generate the  $x_0$ and $x_M$ configurations  according to the
probability distribution $\Psi_0(x_0) \Psi_0(x_M)$, by means of 
a QMC technique at zero temperature, like the PIGS
algorithm.}
\item{To choose  $M$ (number of points of the discrete path from
$x_0$ to $x_M$), so that the parameter $\varepsilon_m = t_m/M$ is
sufficiently small to recover the $\varepsilon_m \to 0$ limit. In practice,
one selects the value of $M$ that makes $\varepsilon_m = t_m/M < \varepsilon_m^*$,
where the parameter $\varepsilon_m^*$ depends on the accuracy of the approximated
action.}
\item{To generate the configurations $\{ x_1, x_2, \ldots ,x_{M-1} \}$,
i.e., the path from $x_0$ to $x_M$, according to the probability
distribution $p_{\text{path}}$}.
\item{To evaluate $O_A(x_0, \ldots, x_M)$ from Eq.
\ref{eq:EstimatorPropagator} and accumulate the estimator of
$C_{AB}(t_c)$ defined in Eq. \ref{eq:CMontecarlo}.}
\end{enumerate}

\subsection{Inversion technique}

Once we have obtained the QMC data for the complex-time correlation
function $C_{AB}(t_c)$, we need to recover the desired spectral function
$S_{AB}(\omega)$ inverting the integral transform
\begin{equation}\label{eq:transform}
C_{AB}(t_c) = \int d\omega \, e^{-i t_c \omega} S_{AB}(\omega) \ .
\end{equation}

Considering that both the function $C_{AB}(t_c)$ and $S_{AB}(\omega)$
are evaluated over a finite set of complex times $\{t_{c \, i}\}$ and
frequencies $\{\omega_j\}$, Eq. \ref{eq:transform} is formally equivalent
to a linear equation
\begin{equation}
y=A\, x    \  ,
\label{eq:directa}
\end{equation}
where the vector $y$ represents the QMC data for the correlation function
$C_{AB}(t_c)$, the vector $x$ the spectral function
$S_{AB}(\omega)$ that we want to obtain, and $A$ is a matrix defined from
the kernel of the integral transform (\ref{eq:transform}) which relates
$C_{AB}(t_c)$ and $S_{AB}(\omega)$. Notice that $C_{AB}(t_c)$ is a complex
function: thus, its real and its imaginary part provide two different rows
of the matrix $A$, both of them real. 

The best least-squares solution to Eq. \ref{eq:directa}  is given by the pseudo-inverse matrix
\begin{equation}
x = A^{T}(A\, A^{T})^{-1}\, y  \ .
\label{eq:reversa}
\end{equation}
In well-posed problems, Eq. \ref{eq:reversa} directly provides useful
solutions. If  $x$ has larger dimensionality than $y$, then the linear equation
in (\ref{eq:directa}) has an infinite number of solutions, and
(\ref{eq:reversa}) provides the one which
minimizes $|x|^2$.  Contrarily, if $x$ has  lower dimensionality than $y$, then no
solution exists and Eq. \ref{eq:reversa} (using the Moore-Penrose pseudoinverse if
$A A^T$ is not full-rank) provides the $x$ vector which minimizes $|y - A x|^2$, i.e., 
a best fit to the $y$ data is obtained.

However, when the eigenvalues of the matrix $A A^T$, which are all positive
or zero, span a range of many orders of magnitude (in the
numerical inversions performed in the present work, eigenvalues of $A
A^{T}$ covering
the range $10^0$-$10^{-20}$ are routinely found), the inversion problem becomes
ill-posed, and the solution $x$ to Eq. \ref{eq:reversa} is extremely sensitive to errors in the
vector $y$.  The ill-posed
nature of the inversion process means that the statistical noise in the
original data for $C_{AB}(t_c)$, that is unavoidable in any QMC calculation, 
is uncontrollably magnified in the inversion
process, resulting in a meaningless solution for the spectral function
$S_{AB}(\omega)$.

In these situations, regularization techniques are useful
to  obtain meaningful solutions to the ill-posed problem.~\cite{key-1} The
basic idea of these methods is to define a well-conditioned linear
operator $C_a$ which depends on a regularization parameter $a>0$
that approaches the pseudo-inverse $A^+ = A^T (A A^T)^{-1}$ in the
limit $a \to 0$. Then, the solution of the original problem can be obtained
as $x = \lim_{a \to 0} C_a y$.

In this work, we have chosen to use the Tikhonov
regularization,~\cite{key-2} in which
\begin{equation}\label{eq:tikhonov}
C_a = A^T (A \, A^T + I a^2)^{-1} \ ,
\end{equation}
where $I$ is the identity matrix. 
Thanks to Tikhonov regularization, the solution
$x$ of the problem is much less sensitive to errors in the initial
vector $y$.  
On the other hand, the regularization procedure introduces a
bias in the estimation of $x$. 
The goal is however to keep the regularization parameter $a$ as small as
possible yo avoid introducing unwanted artifacts in the reconstructed
solution.

In practice, the choice of the regularization parameter must avoid two different
problems. If the regularization parameter $a$ is too small, the solution is 
unstable and  similar QMC data for the correlation function lead to different spectral
functions. If $a$ is too large, systematic effects start to appear in the
solution.  These effects can
be controlled verifying that the correlation function obtained applying the direct integral
transform   (Eq. \ref{eq:transform}) to the given solution for the spectral function 
is in agreement with the
starting QMC data for $C_{AB}(t_c)$ (see Appendix B for
additional information). Monte Carlo data of higher quality allow 
for smaller values of
the regularization parameter and thus they are crucial for a
satisfactory direct inversion.

Focusing on the dynamic structure factor, 
the physical solution must verify $x_{i}\geq 0$ for every component
of $x$   since $S(q,\omega) \geq 0$ . 
We introduce this
requirement explicitly in the construction of the solution, making
use of a square diagonal matrix $Q=\text{Diag}(q_{1},\ldots,q_{N})$, where
each of the $q_{i}$ is to be understood as a factor (which we restrict
to be either $0$ or $1$) that will multiply explicitly the component
$x_{i}$ of the vector solution $x$. The new solution, that can be
written formally as 
\begin{equation}
x=Q\, A^{T}(A\, Q\, A^{T})^{-1}\, y   \ ,
\label{eq:ambQs}
\end{equation}
satisfies by construction both $x_{i}=0$ if $q_{i}=0$ and $y=A\, x$, 
irrespective of $Q$. The regularization procedure can be performed
in this case by simply making the substitution 
$A\, Q\, A^{T}\rightarrow A\, Q\, A^{T}+I\: a^{2}$.
We use Eq. \ref{eq:ambQs} as a means of imposing the positiveness
of $S(q,\omega)$. In order to do so, we set an iterative procedure, starting
with $Q=\text{Diag}(q_{1}=1,\ldots,q_{N}=1)$, using the regularized version
of Eq. \ref{eq:ambQs}, to obtain the vector solution $x$, and we
set $q_{i}=0$ for all components $x_{i}<0$ and form a new $Q$ matrix
which contains more zeroes in the diagonal than the previous one.
Inserting the new $Q$ back in Eq. \ref{eq:ambQs}, a new solution is
obtained. The procedure is repeated until no negative components are
present, and we end up with a regularized, positive solution to the
inversion problem.

\section{Results}

The formalism developed in Sec. II has been applied to the calculation of
the density-density correlation function in complex time,
\begin{equation}
S(q,t_c)= \langle \Psi_0 | e^{it_c \hat{H} / \hbar} \hat{\rho}_q  e^{-it_c \hat{H} / \hbar} 
\hat{\rho}_{-q} | \Psi_0 \rangle \ ,
\label{sqtc}
\end{equation}
with the density-fluctuation operator $\hat{\rho}_q = \sum_{i=1}^{N} e^{i
{\bf q} \cdot {\bf r}_i}$ and complex time $t_c=t_m e^{-i \delta}$. The
reliability of the method has been checked in two model problems which can
be easily solved: a particle in a one-dimensional harmonic potential (HP),
$V(x)=x^2/2$,  and a particle in a one-dimensional anharmonic potential (AP),
$V(x)=x^4/4$. We work in units where $\hbar=m=1$. The ground-state wave
function $\Psi_0$ (\ref{sqtc}) is obtained using the PIGS algorithm with the high-order Chin
action.~\cite{chin,Rota10}

\begin{figure}
\includegraphics[width=0.5\textwidth,angle=-90]{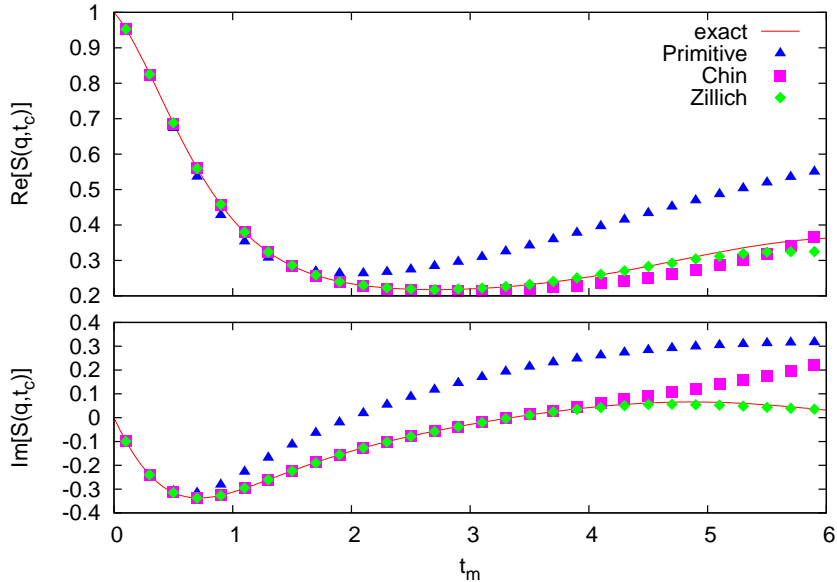}
\caption{(Color online) Real (top) and imaginary (bottom) parts of
$S(q,t_c)$ for HP, with $q=1.5$ and $\delta=\pi/9$, as a function of
$t_m=|t_c|$.
The line stands for the exact result (\ref{sqtc_hp}) and the points to
different approximations for the action. Triangles, primitive action;
squares, Chin action;~\cite{chin} diamonds, Zillich action.~\cite
{Zillich10}  }
\label{fig:action}
\end{figure}

As commented in Sec. II, a relevant aspect that makes the
calculation in complex time be more accurate is to use high-order actions in
the evaluation of Eq. \ref{sqtc}. We need to work with as few number of
beads $M$ as possible to reach the maximum accessible time. In Fig.
\ref{fig:action}, we show results for the real and imaginary parts of
$S(q,t_c)$ for the HP as a function of $t_m$. The results correspond
to $q=1.5$ and $\delta=\pi/9$. The line stands for the HP exact
result,~\cite{Lovesey}
\begin{equation}
S(q,t_c)= \exp \left[ \frac{q^2}{2} \left( e^{-i t_c} -1 \right) \right]  \ .
\label{sqtc_hp}
\end{equation} 
In the figure, we compare the exact function (\ref{sqtc_hp}) with our QMC
results obtained with a single bead, $M=1$, using different approximations
for the actions employed  in the evaluation of $S(q,t_c)$. As expected, the
PA is 
only accurate at very short times. 
If we consider QMC results for $S(q,t_c)$ with a relative error 
of 0.4\%, we notice that these are in agreement with the exact result 
for $t_m \lesssim 0.3$ and depart significantly of the exact 
result al larger time. Therefore, the PA is not a good choice because we
would need a large number of beads to 
span the full time range. The results
are significantly better if one uses high-order actions. In the figure, we
show estimations of the real and imaginary parts using the Chin
action~\cite{chin} and a sixth-order expansion reported by Zillich
\textit{et al.}~\cite{Zillich10} Comparing numerical results of $S(q,t_c)$, 
with the same precision as before, we notice that the    
Chin action reproduces the exact results
up to $t_m \simeq 2$. However, the Chin
action is in general not appropriate because of the divergence terms
derived from the double commutator (notice that for the HP this divergence
is reduced because this contribution produces a renormalization of the
oscillator frequency). The
best result is obtained using the sixth-order
approximation.~\cite{Zillich10} 
This action is able to account for the exact data up to $t_m \simeq 3.5$ and
with the added benefit of not requiring double-commutator terms since it is based
on extrapolations of PA actions with different time steps. Therefore, we
have selected this action as the best option for this complex-time
estimation.


A second step in our methodology is the estimation of $\varepsilon_m^*$
(see Sec. II) which determines the maximum time $t_m$ that can be covered
with a single bead, with no significant bias coming from the small-time approximation
of the action. This estimation is performed by studying the convergence
of $S(q,t_c)$, with $t_m = |t_c|$ fixed, for small values of $\varepsilon_m
= t_m/M$. To perform this analysis, we have selected $\delta=\pi/2$
(imaginary time). Using a different
value of $\delta$, the statistical error of $S(q,t_c)$ tends to increase
largely with the number of beads $M$ because the phase of the estimator of
$S(q,t_c)$ is proportional to $\cos \delta$ (see Appendix A), and it
is not possible to give precise estimates in the limit of small
$\varepsilon_m$. 

\begin{figure}
\includegraphics[width=0.5\textwidth,angle=-90]{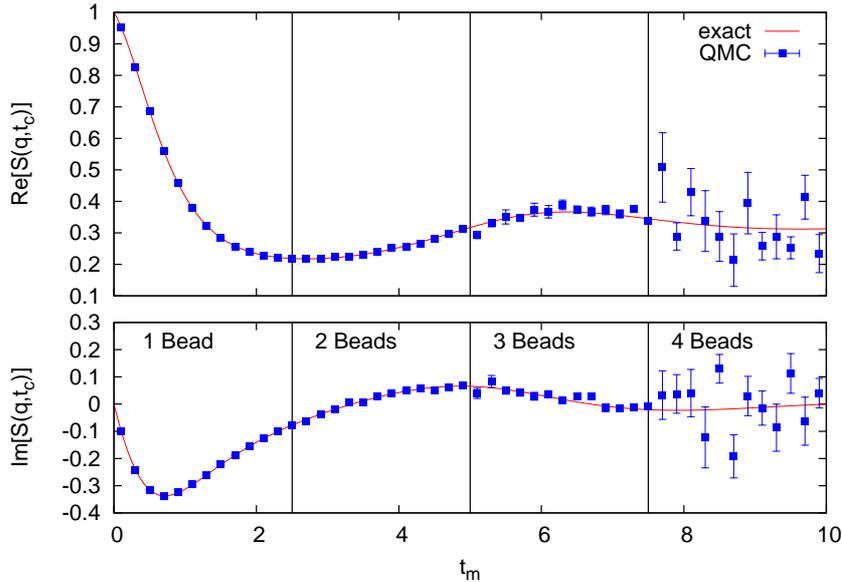}
\caption{(Color online) Real and imaginary parts of $S(q,t_c)$ for HP, 
with $\delta=\pi/9$, as a function of $t_m$. The line is the exact
expression
(\ref{sqtc_hp}) and the points correspond to our QMC results. The vertical lines
separate the results obtained with different number of beads $M$. 
Where not shown, error bars are smaller than the symbol size.}
\label{fig:empalme}
\end{figure}

With the estimation of the accuracy of the action 
(for HP, we get $\varepsilon_m= 2.5$),
one can easily determine the number of complex-time beads required in the
calculation of $S(q,t_c)$ at any $t_c$: $M$ is the minimum integer for which the
condition $|t_c|/M < \varepsilon_m^*$ is satisfied. Accordingly, 
the whole range of times $t_m=|t_c|$ is divided in different regions
where $S(q,t_c)$ is estimated with a different number of beads. In
practice, $M=1$ for $t_m \in [0,\varepsilon_m^*]$, $M=2$ for $t_m \in
[\varepsilon_m^*,2 \varepsilon_m^*]$, and so on. The results obtained with
this splitting are reported in Fig. \ref{fig:empalme} for the HP and
$\delta=\pi/9$. In the figure, the vertical lines separate the different
intervals $[(M-1) \varepsilon_m^*, M \varepsilon_m^*]$ where $S(q,t_c)$ is
calculated with the same number of beads $M$. The trends observed in this
particular case are quite general. The results obtained are statistically
reliable up to a maximum time $t_m$ which decreases when the phase $\delta$
is reduced. This feature directly implies that the maximum number of beads
producing sound results is also reduced when approaching the real axis. In
general, the number of beads is small but the high accuracy of the action
used in the calculation makes the total covered time be quite large. In the
case shown in Fig. \ref{fig:empalme}, one can see that our QMC estimation
is satisfactory up to $M_{\text{max}}=3$, with a total time 
$t_m= M_{\text{max}} \varepsilon_m^* =
7.5$. The results with $M=4$ are spread around the exact function but with
too large error bars to be used in the subsequent transform to the
dynamic structure function $S(q,\omega)$.

\begin{figure}

\includegraphics[width=0.4\textwidth,angle=0]{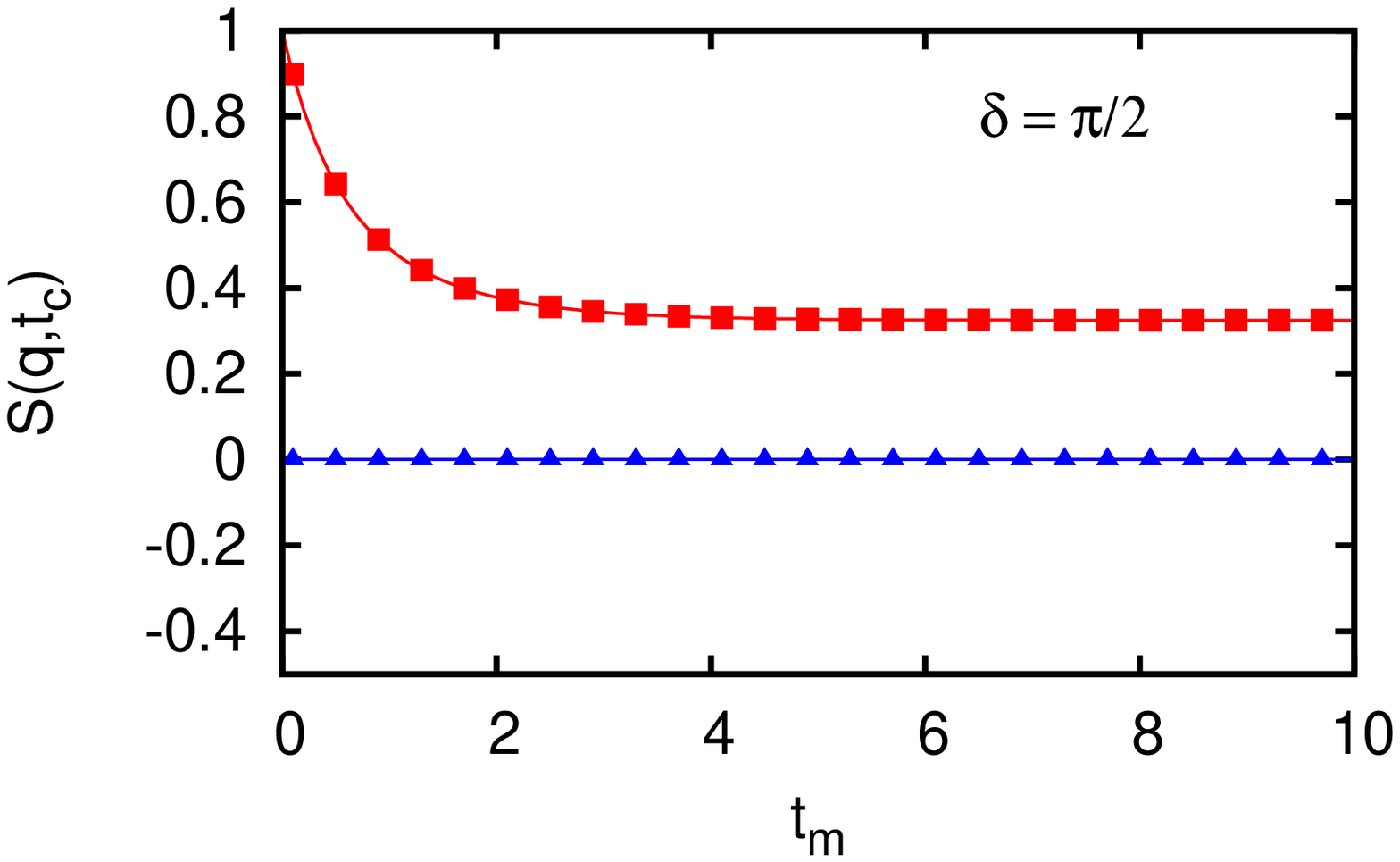} \\
\includegraphics[width=0.4\textwidth,angle=0]{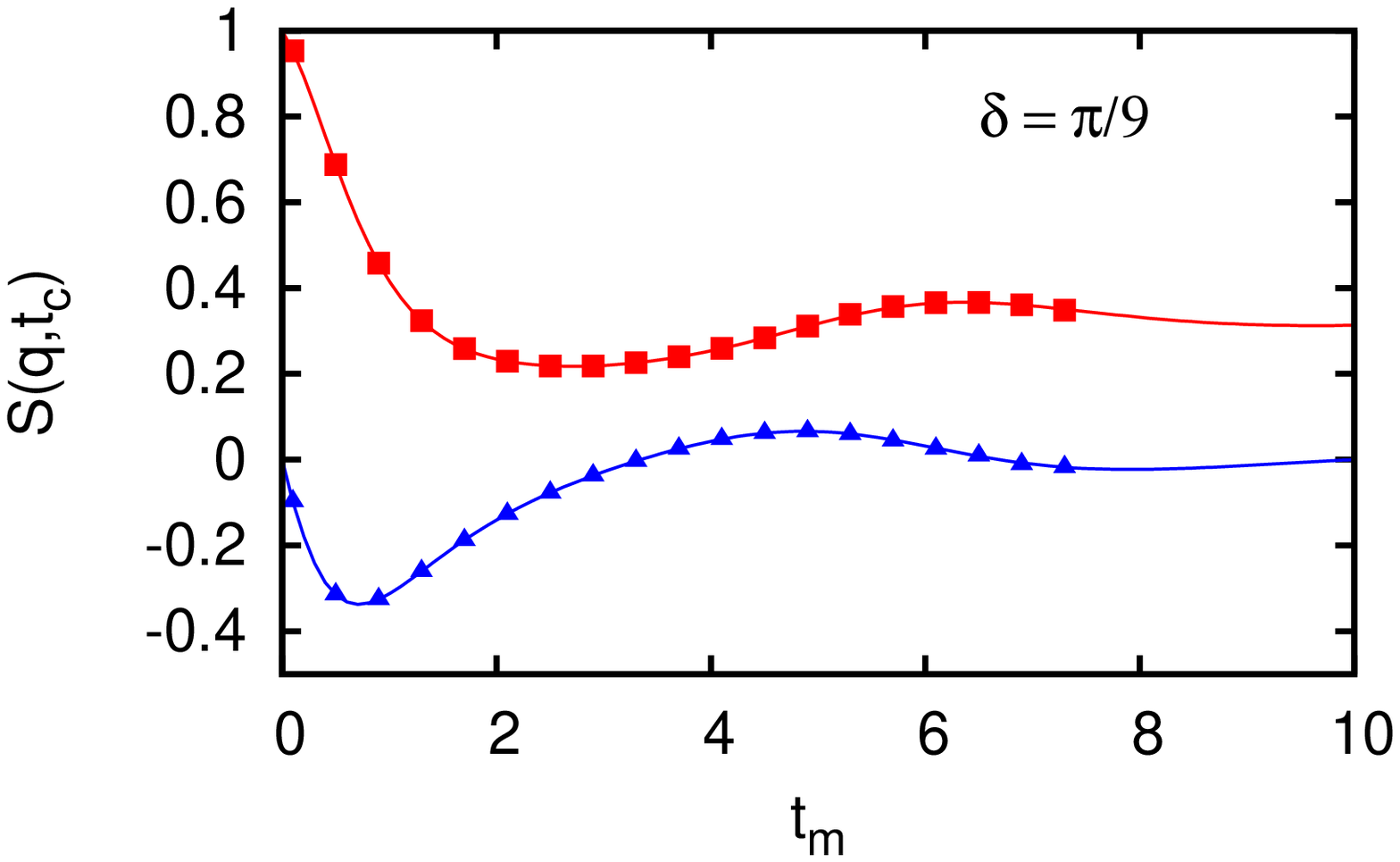}\\
\includegraphics[width=0.4\textwidth,angle=0]{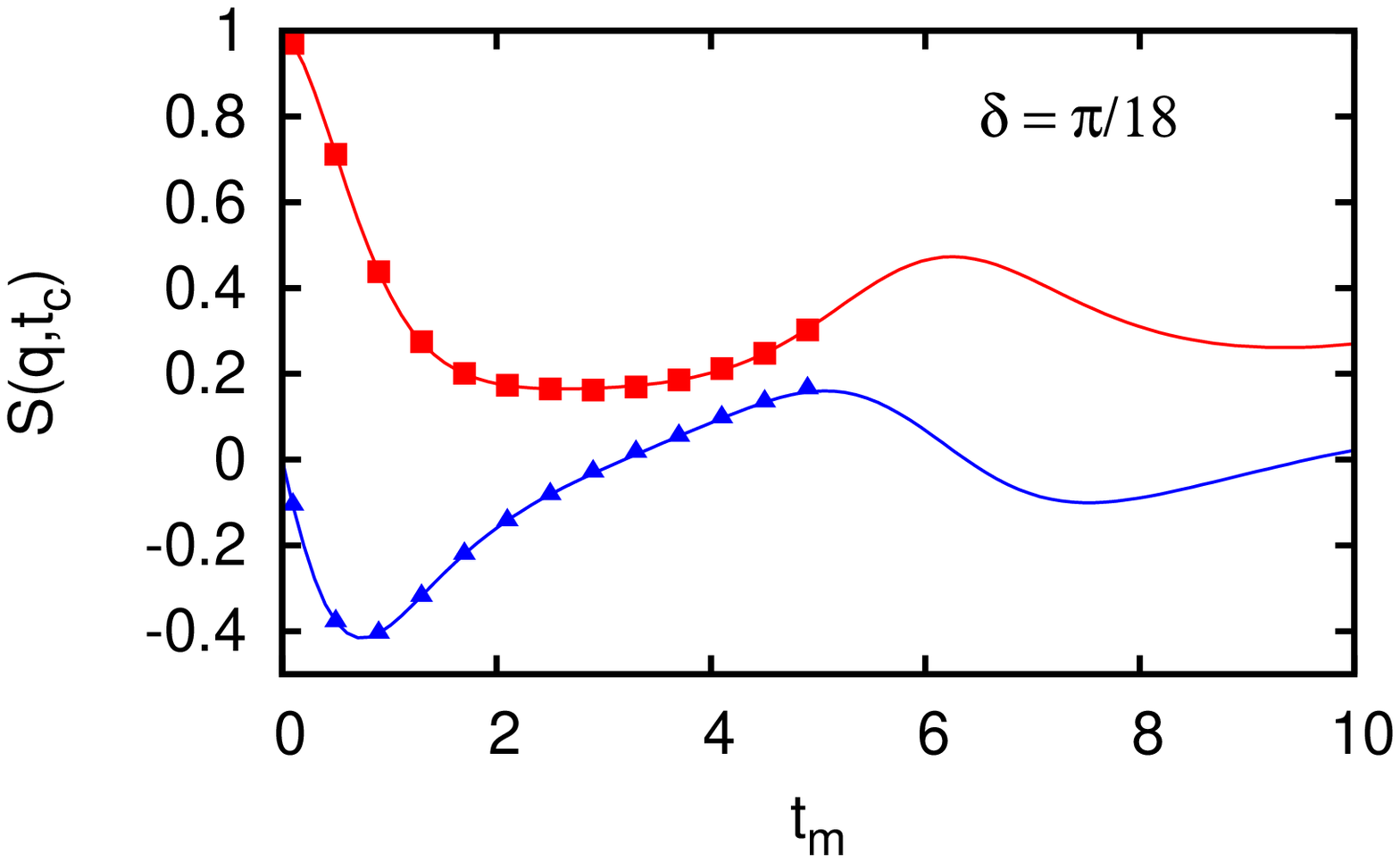}

\caption{(Color online) Real and imaginary parts of $S(q,t_c)$ for HP 
as a function of $t_m$. The upper and lower lines in each panel correspond to the 
real and imaginary parts of the exact result
(Eq. \ref{sqtc_hp}), respectively. 
The symbols correspond to our QMC results 
(squares: real part; triangles: imaginary part). Each panel corresponds 
to the calculation of $S(q,t_c)$ for different values of the phase $\delta$ 
of the complex time. Error bars are smaller than the symbol size.}
\label{fig:sqtchp_full}
\end{figure}

In Fig. \ref{fig:sqtchp_full}, we show QMC results of the complex function
$S(q,t_c)$  for the HP and different values of the phase $\delta$, in comparison
with the exact function (\ref{sqtc_hp}).  When approaching the
real axis, i.e. when $\delta$ decreases, both the real and imaginary
parts show an increase of their oscillatory behavior (notice that for HP,
the exact $S(q,t)$ for real time is periodic), but the maximum reachable
value $t_m$ decreases. Therefore, there is a compromise
between lowering $\delta$ as much as possible and reaching times as large as
possible. Our results show that the optimal phase for a posterior transform
to the frequency domain is within the range $[\pi/18,\pi/9]$.

\begin{figure}

\includegraphics[width=0.4\textwidth,angle=0]{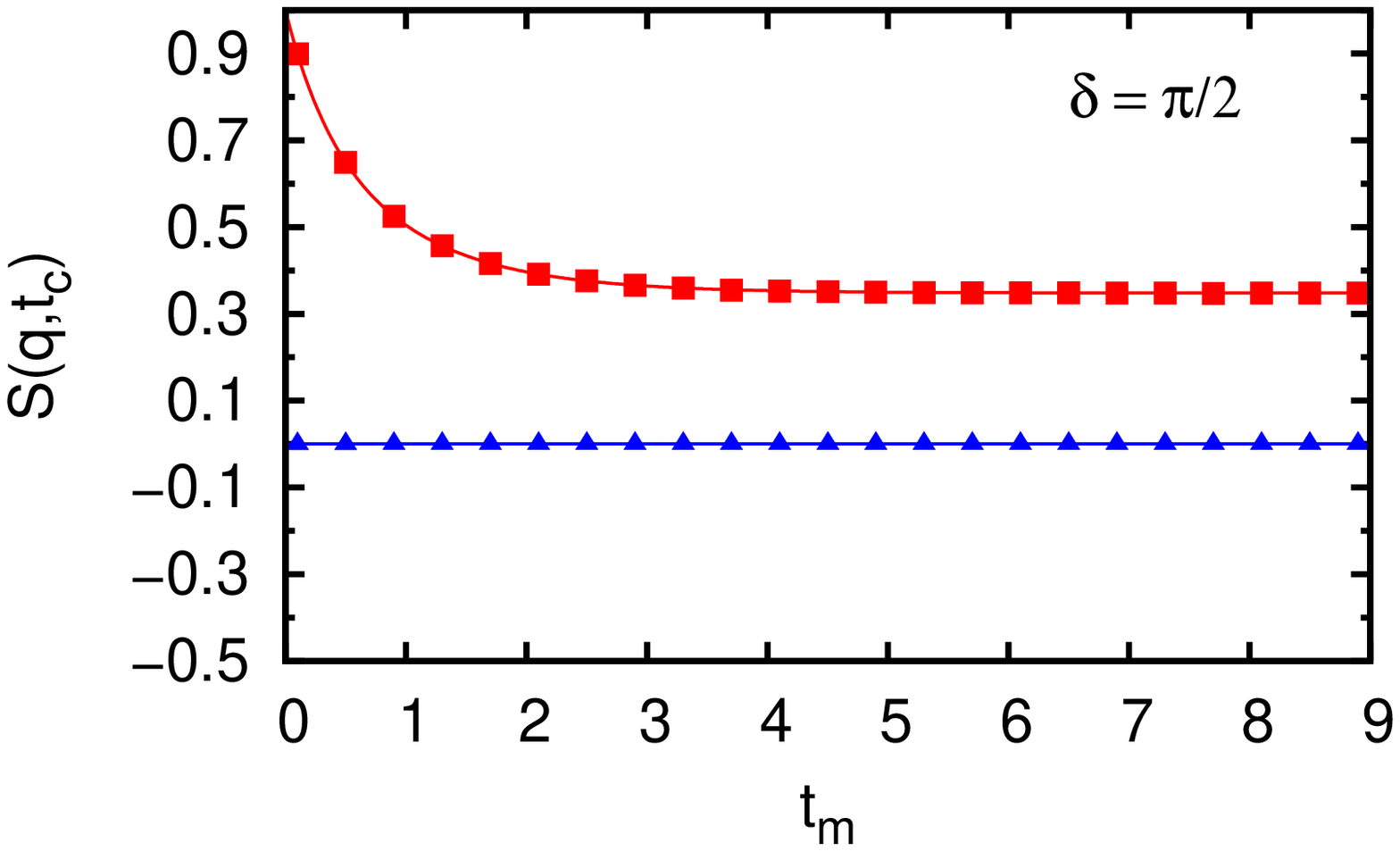} \\
\includegraphics[width=0.4\textwidth,angle=0]{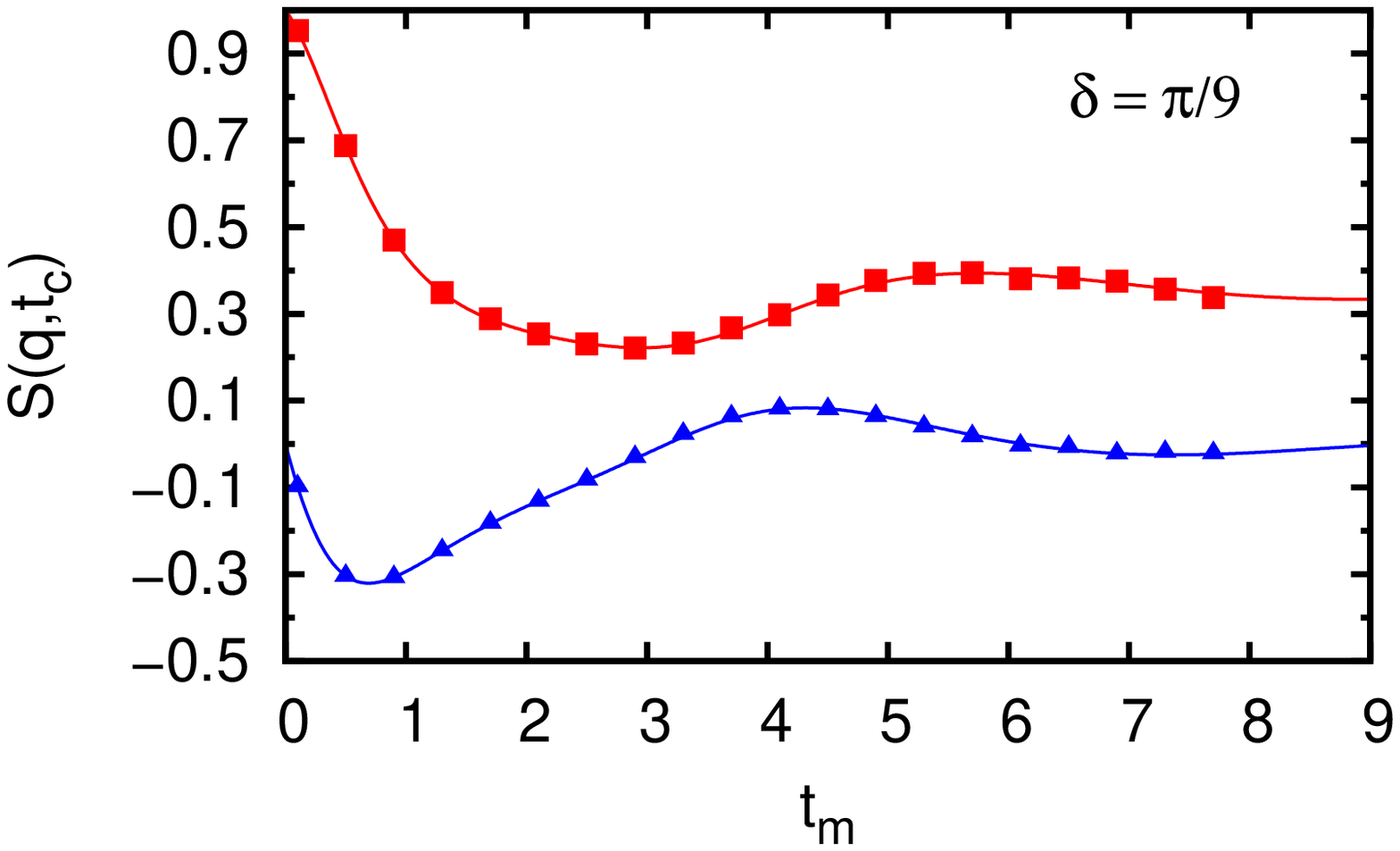}\\
\includegraphics[width=0.4\textwidth,angle=0]{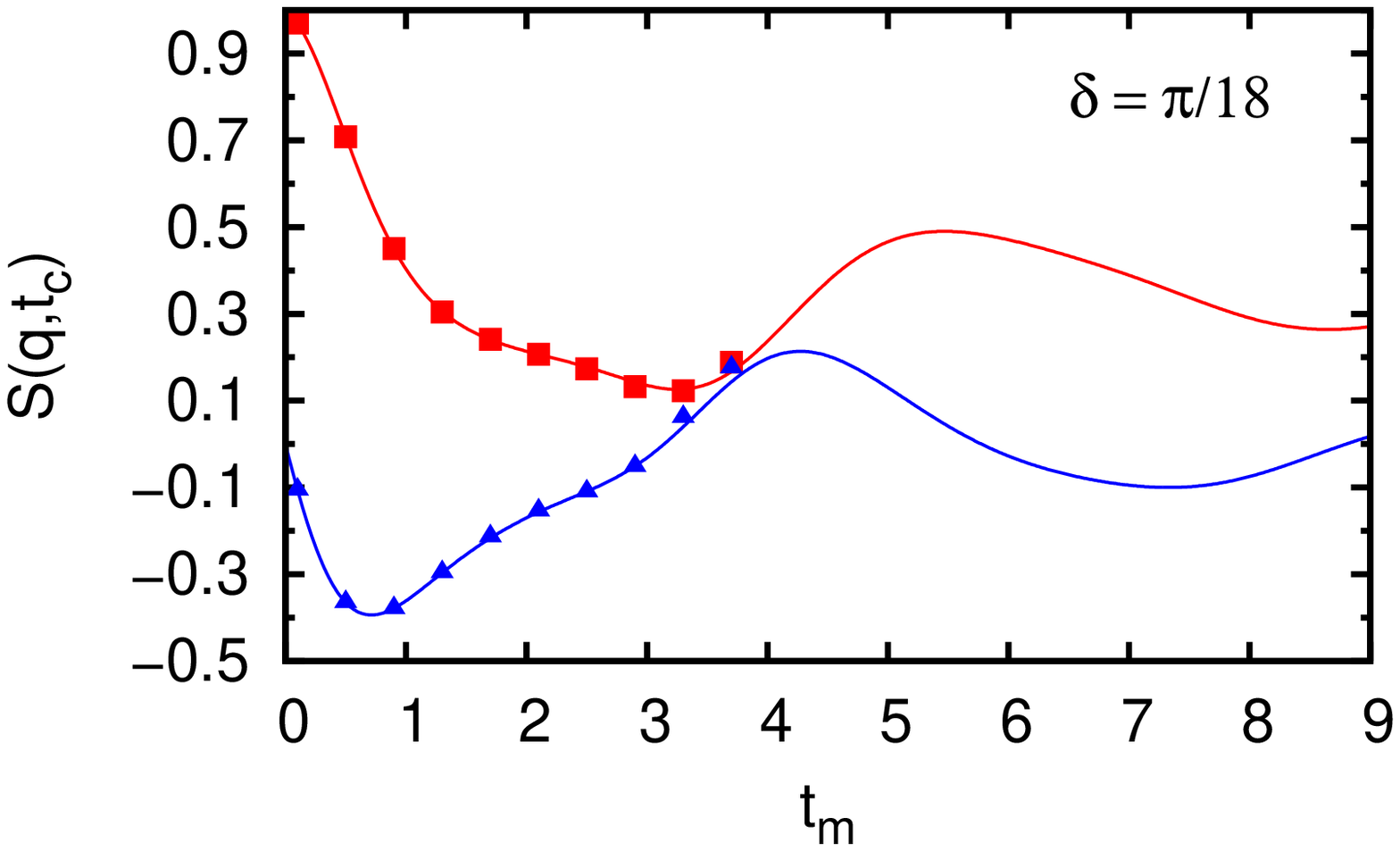}

\caption{
(Color online) Real and imaginary parts of $S(q,t_c)$ for AP 
as a function of $t_m$. The upper and lower lines in each panel correspond to the 
real and imaginary parts of the exact result, respectively. 
The symbols correspond to our QMC results 
(squares: real part; triangles: imaginary part). Each panel corresponds 
to the calculation of $S(q,t_c)$ for different values of the phase $\delta$ 
of the complex time. Error bars are smaller than the symbol size.}
\label{fig:sqtcap_full}
\end{figure}

Proceeding in a similar way we have applied our method to the study of the
correlation function for a particle in an AP. The results for the
real and imaginary parts of $S(q,t_c)$  are shown in Fig.
\ref{fig:sqtcap_full} for different values of the phase $\delta$ ranging
from $\pi/2$ (imaginary time) down to $\pi/36$. Our Monte Carlo results are
compared with exact ones obtained by numerical integration over the
eigenstates of the Hamiltonian (differently to the HP case, an analytical form
for the $S(q,t_c)$ of the AP is not known). The QMC estimation of the 
complex-time correlation functions
 shows similar accuracy to the one achieved for the HP
case. Similarly to HP, we recover for the AP the exact results up to a 
maximum value of the modulus of
the complex time $t_m$. Beyond this value, which decreases with $\delta$, the
statistical errors grow significantly, making any estimation of
$S(q,t_c)$ not reliable. Again, a good compromise between statistical fluctuations and
approaching the real axis as close as possible locates the optimal values of
the phase in the same range than in the HP case, $\delta \in [\pi/18,\pi/9]$. 

Once we have found the working window,  the
next step is to make the inversion from complex-time to energies.
Our goal is to calculate the dynamic response $S(q,\omega)$ and compare the
results with the exact function for both the HP and
AP. To this end, we have applied the inversion technique described in the
previous Section. A preliminary point is to know up to which extent the
inversion procedure can influence the results in the energy domain. In the
case of purely imaginary-time data, several inversion methods have been
used,~\cite{Boninsegni96,Sandwik98,Mishchenko00,Vitali10} the majority of
them being of stochastic nature. This inverse Laplace transform  
is normally mapped to a multidimensional optimization problem. The
ill-posed nature of this inversion can lead to results that
can depend on the method employed.

\begin{figure}

\includegraphics[width=0.45\textwidth,angle=0]{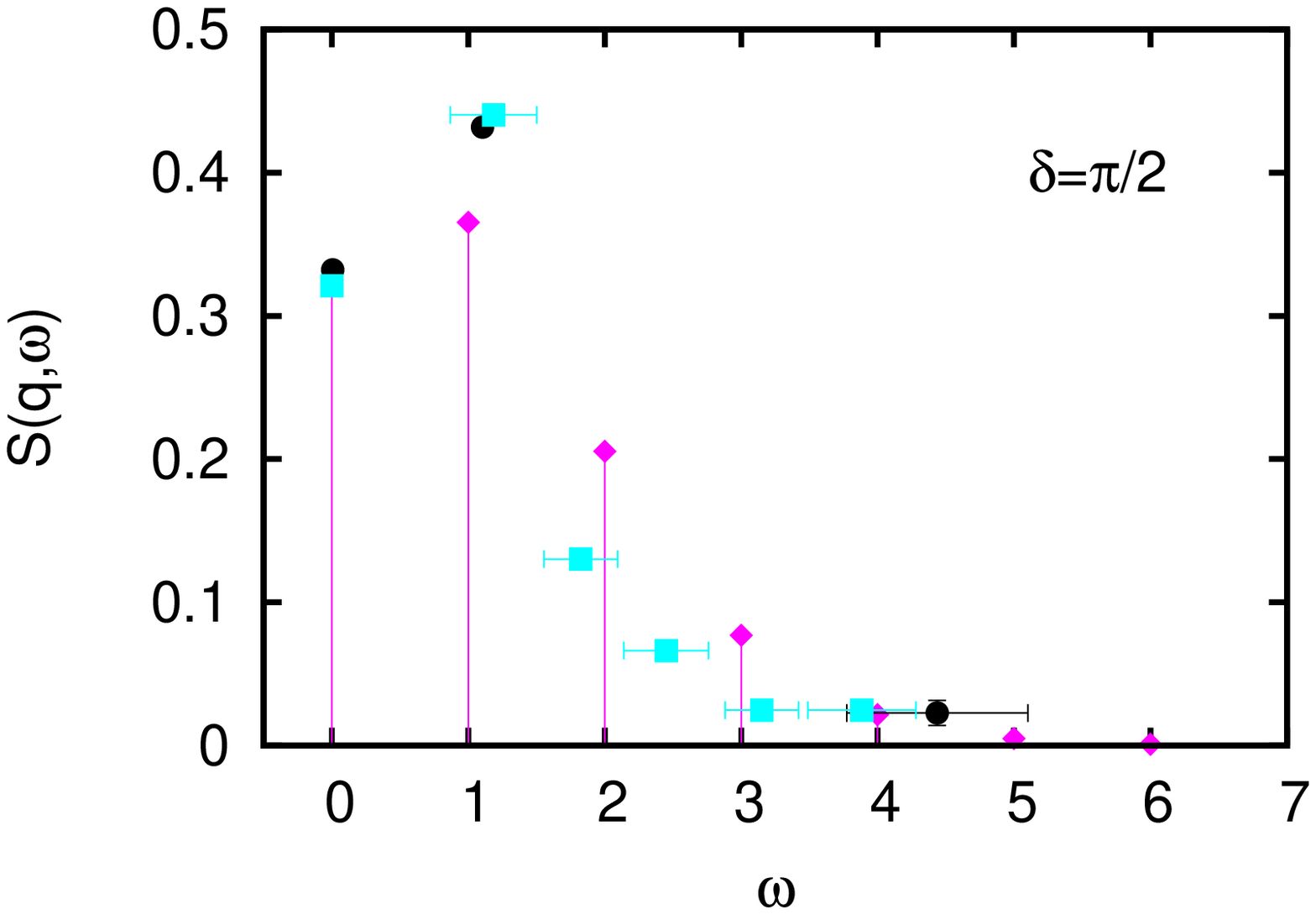}
\includegraphics[width=0.45\textwidth,angle=0]{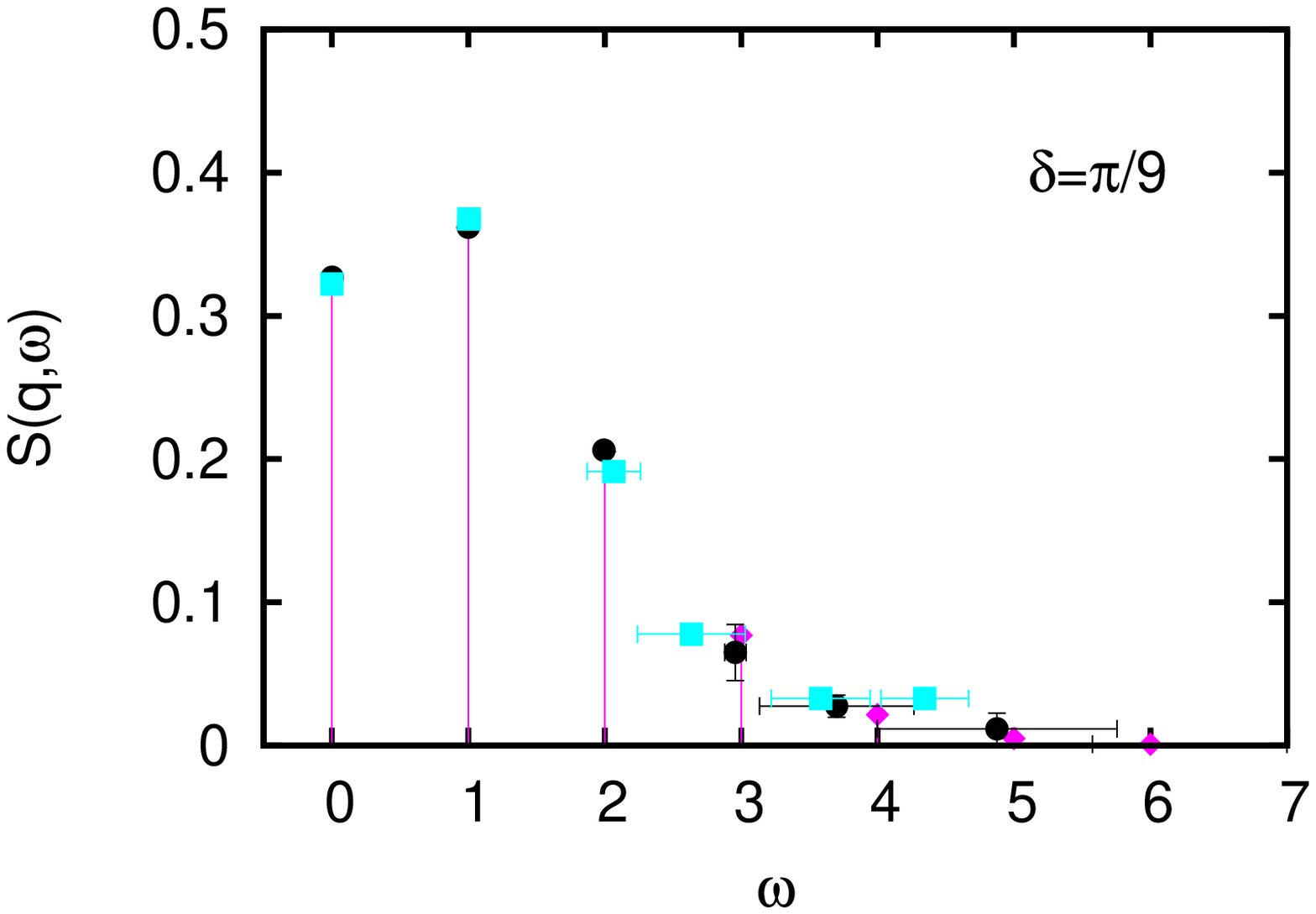}

\caption{(Color online) Dynamic structure function $S(q,\omega)$ for the
HP at $q=1.5$. Diamonds correspond to the exact values and 
and circles and squares with errorbars to
the results derived from the QMC results for $S(q,t_c)$. The circles are
obtained using the method described in Sec. II and the squares 
using a standard simulated annealing schedule.
Left panel: imaginary time ($\delta=\pi/2$). Right panel:
complex time ($\delta=\pi/9$).   }
\label{fig:sannealing}
\end{figure}

In Fig. \ref{fig:sannealing}, we compare results obtained for $S(q,\omega)$
in the HP problem using the inversion method discussed in the previous Section
and a standard simulated annealing algorithm. In the figure, the exact
result~\cite{Lovesey}
\begin{equation}
S(q,\omega)= e^{-q^2/2} \, \sum_{n=0}^{\infty} \frac{1}{2^n n!} q^{2n}
\delta(n-\omega)
\label{sqwexhp}
\end{equation} 
is also plot with vertical lines.
This comparison is made for two cases:
imaginary-time data ($\delta=\pi/2$) and complex-time results with
$\delta=\pi/9$. As it has been commented, the inversion from
imaginary time  to the frequency domain is an ill-posed problem and thus the results can show
differences depending on the selected method. This is shown in Fig.
\ref{fig:sannealing} (left panel): the inversion obtained from the stochastic simulated
annealing method and the one discussed in Sec. II 
produce slightly different predictions for the higher transition lines, while they
both agree on the first and second peaks, although the latter has a total
strength that is $\sim 15$\% off from the exact value in both cases. None of
the high transition lines is well reproduced by any of the two models. 
 In the same figure (right panel),
we compare the results from both inversion methods for $\delta=\pi/9$. In this case, the
inversion works on complex-time data which shows a richer structure. This
significantly reduces the ill-posed character of the inversion and thus the
results obtained with both methods look much more similar than in
the $\delta=\pi/2$ case. Our results show that the three main peaks are
well reproduced and the fourth one is approximated, slightly better using
the non stochastic method which has been computed averaging over a larger data
set.

\begin{figure}

\includegraphics[width=0.4\textwidth,angle=0]{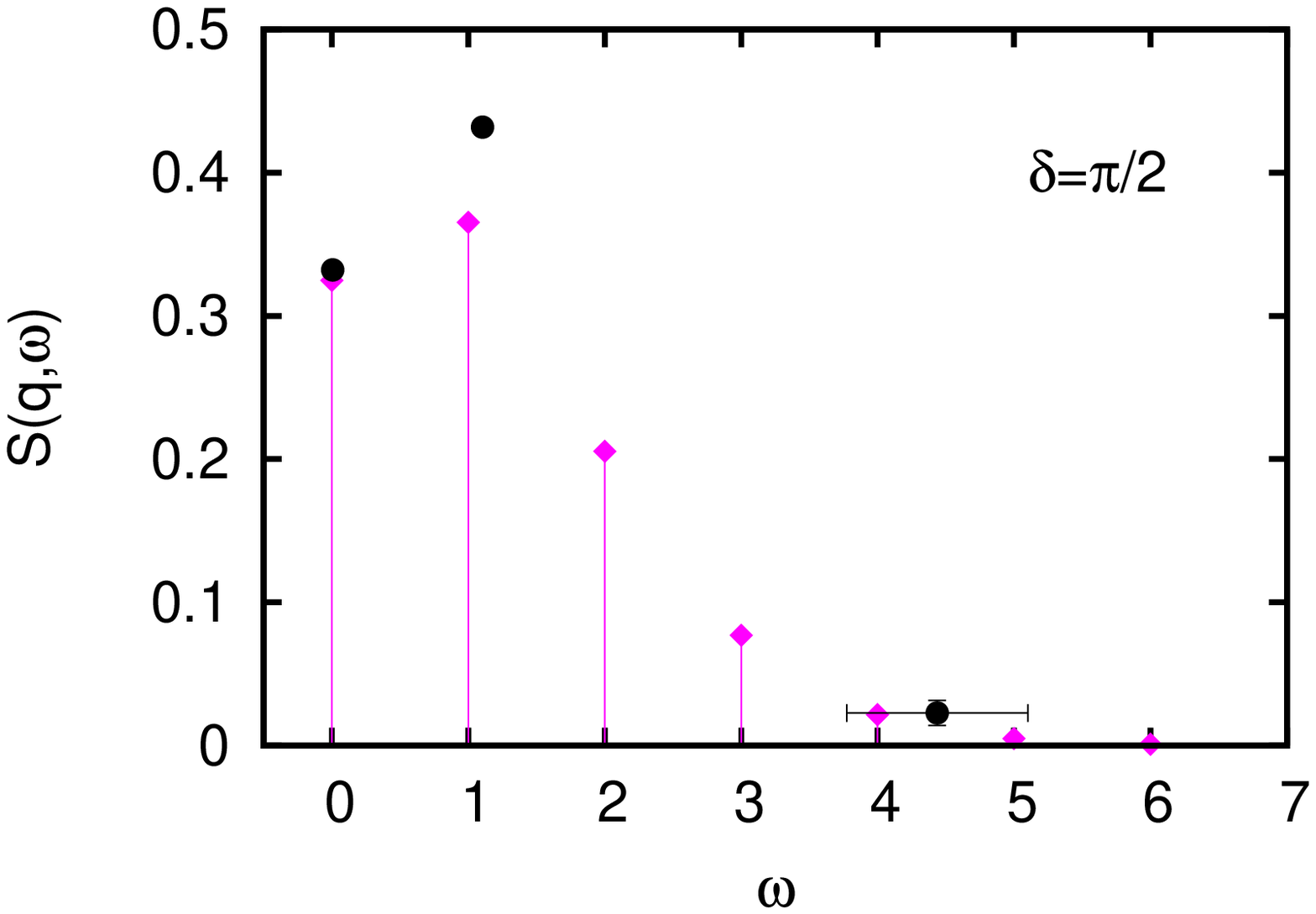}
\\
\includegraphics[width=0.4\textwidth,angle=0]{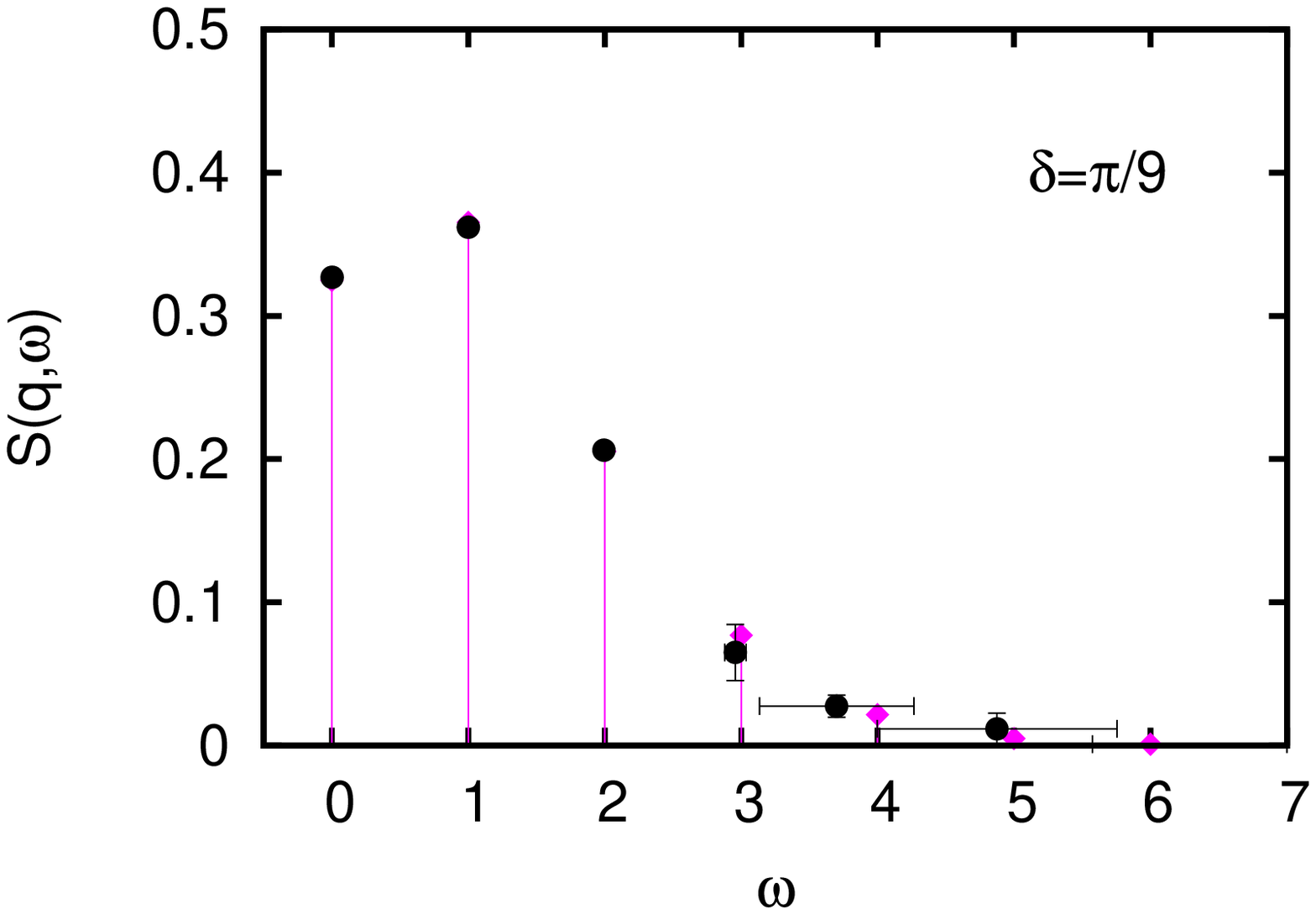} \\
\includegraphics[width=0.4\textwidth,angle=0]{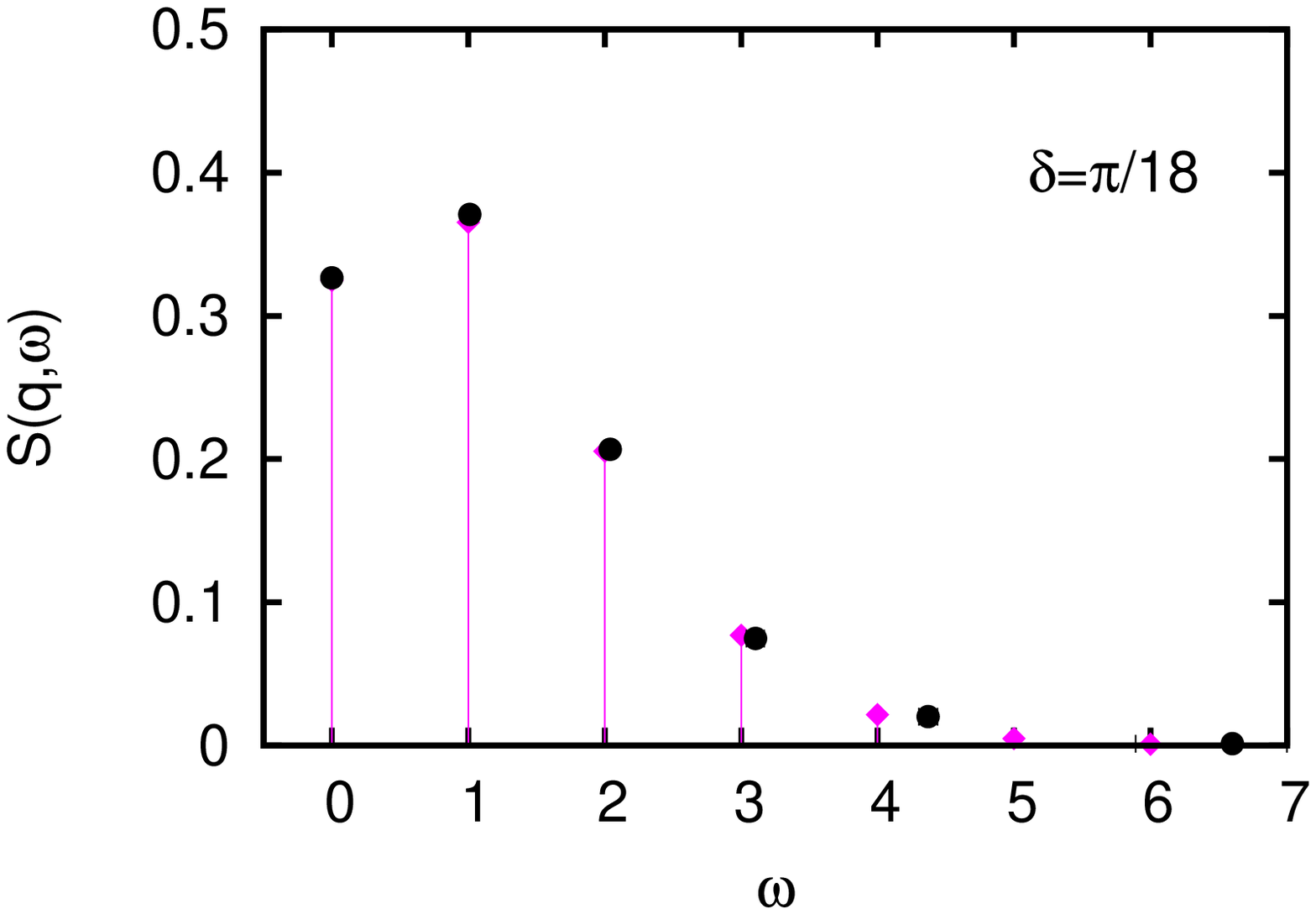}

\caption{(Color online) Dynamic structure function $S(q,\omega)$ for the
HP at $q=1.5$. Diamonds correspond to the exact values (\ref{sqwexhp}) 
and circles with errorbars to
the results derived from the QMC results for $S(q,t_c)$. The inversion
uses complex-time data calculated at the phases $\delta$ reported in each
panel.   }
\label{fig:HO_q15}
\end{figure}

The results that we have obtained for the HP dynamic response at $q=1.5$ are reported in Fig.
\ref{fig:HO_q15}. The different panels contain reconstructions from
imaginary-time data ($\delta=\pi/2$) and complex-time correlation factors
estimated at decreasing values of the phase, down to $\delta=\pi/18$. At  
$\delta=\pi/2$
we recover the first peak (energy and strength) and approximate the
second one. In other words, only the lowest-energy mode is accurately reproduced. It
is worth noticing that this is the overall trend observed in
transformations from purely imaginary-time data. By 
progressively introducing a real component in the correlation factor, i.e., by decreasing
the phase $\delta$, the quality of the dynamic response improves
significantly. As one can see, for $\delta=\pi/9$ one gets the first four
modes with their respective strengths in nice agreement with the exact values.
By reducing even more the phase down to $\delta=\pi/18$ we are able to reduce
the variance of the data but no additional (higher) energies are resolved.
Notice, however, that the strength of the peaks beyond the first four ones
is much smaller.

\begin{figure}

\includegraphics[width=0.4\textwidth,angle=0]{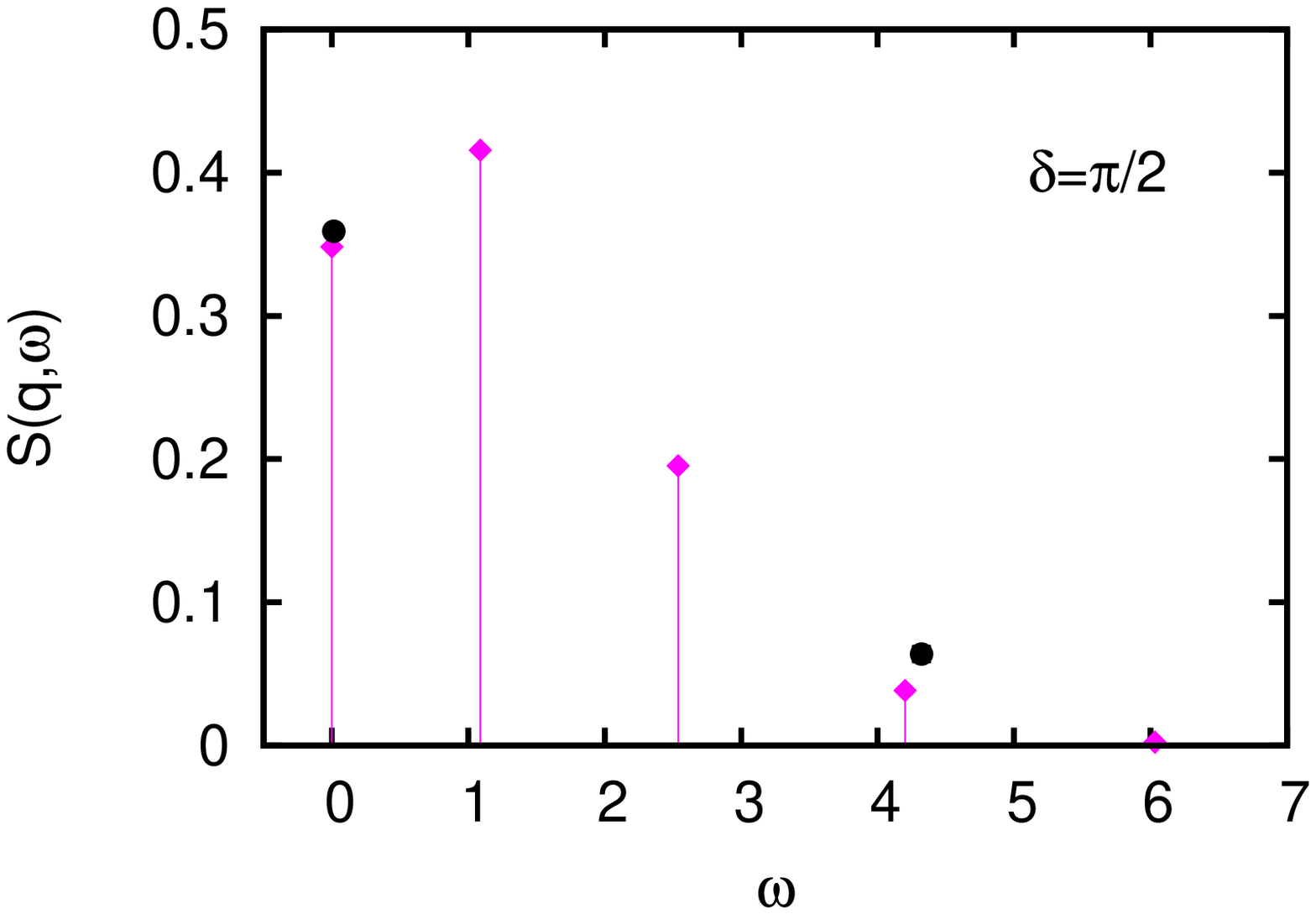}
\\
\includegraphics[width=0.4\textwidth,angle=0]{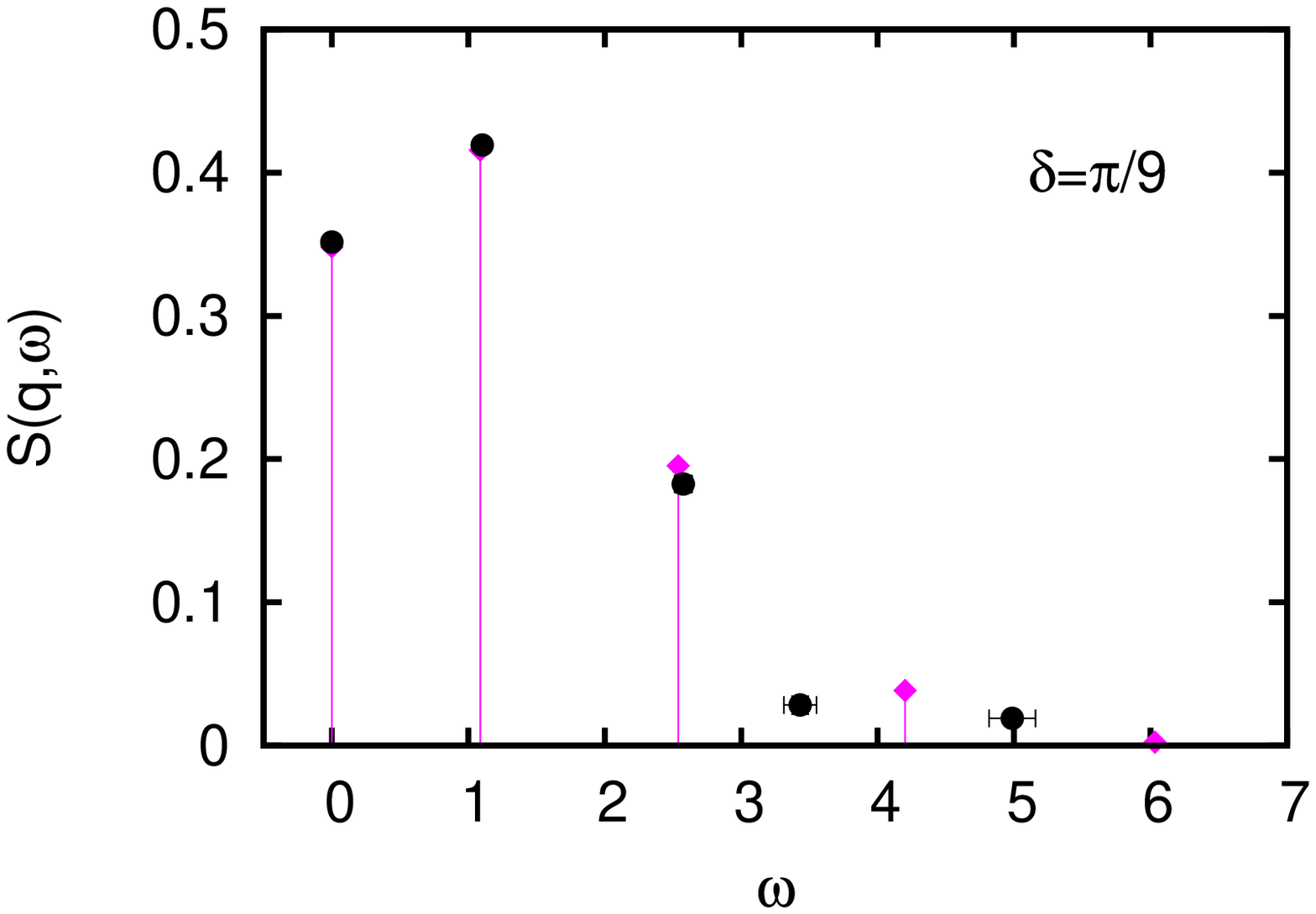} \\
\includegraphics[width=0.4\textwidth,angle=0]{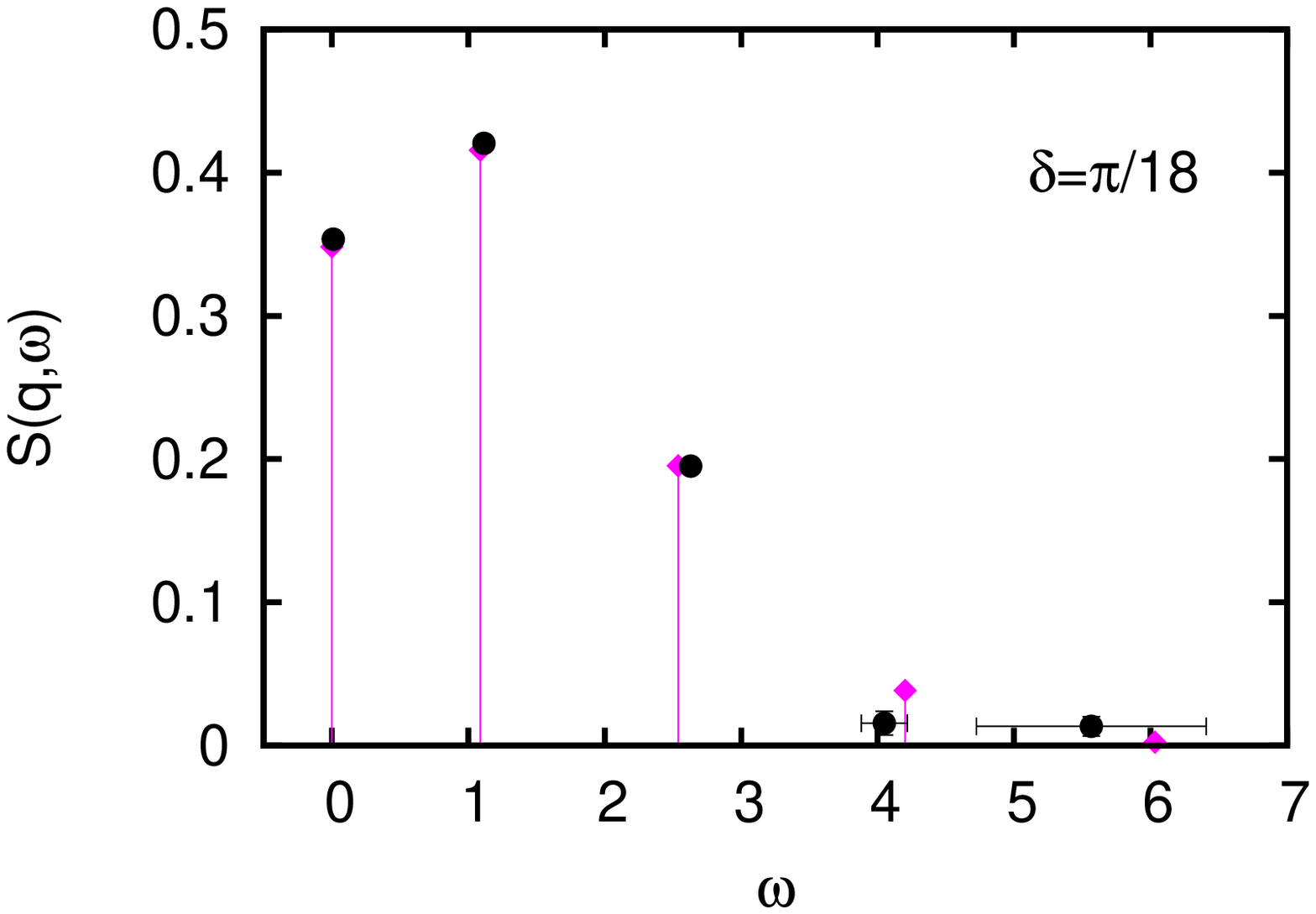}

\caption{(Color online) Dynamic structure function $S(q,\omega)$ for the
AP at $q=1.5$. Diamonds correspond to the exact values and circles with errorbars to
the results derived from the QMC results for $S(q,t_c)$. The inversion
uses complex-time data calculated at the phases $\delta$ reported in each
panel.   }
\label{fig:AP_q15}
\end{figure}

The same analysis has been carried out for the AP. Our results 
of the dynamic structure functions are contained in Fig.
\ref{fig:AP_q15}. With imaginary-time data, we are able to reproduce only the first
peak. By decreasing the phase $\delta$ the dynamic response improves
progressively. At $\delta=\pi/9$, the three main modes and their respective
strengths are in close agreement with the exact results. For the smallest
value $\delta=\pi/18$, we can even resolve the fourth mode whose strength
is already quite small. Again, the gain of working with complex-time
correlation factors becomes evident.

\begin{figure}

\includegraphics[width=0.45\textwidth,angle=0]{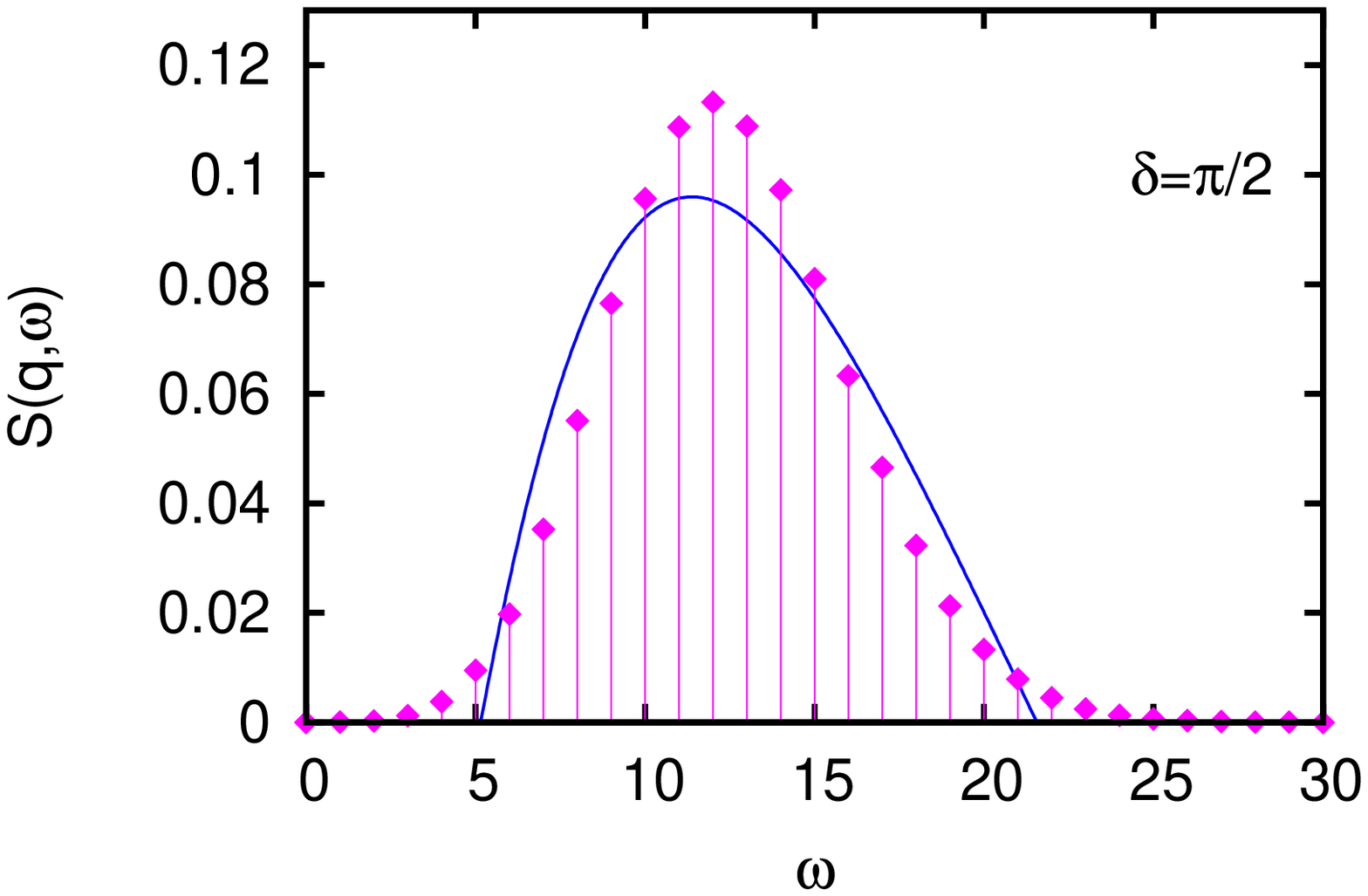}
\includegraphics[width=0.45\textwidth,angle=0]{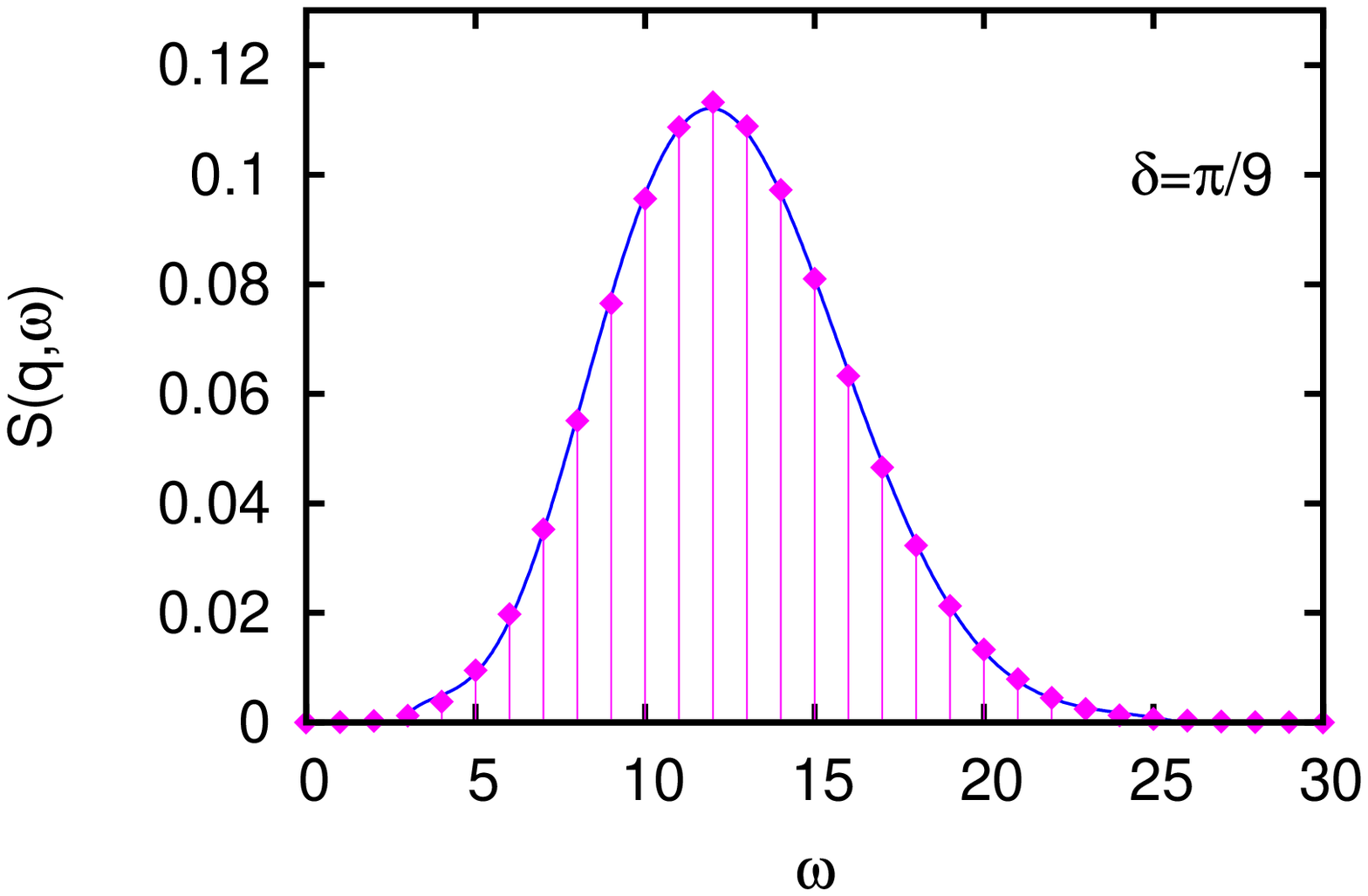}

\caption{(Color online) Dynamic structure function $S(q,\omega)$ for the
HP at $q=5$. Diamonds correspond to the exact values (\ref{sqwexhp})
and the curve corresponds to the results derived from our QMC results.
 Left panel: imaginary time ($\delta=\pi/2$). 
Right panel: complex time ($\delta=\pi/9$).}    
\label{fig:HO_q5}
\end{figure}

\begin{figure}

\includegraphics[width=0.45\textwidth,angle=0]{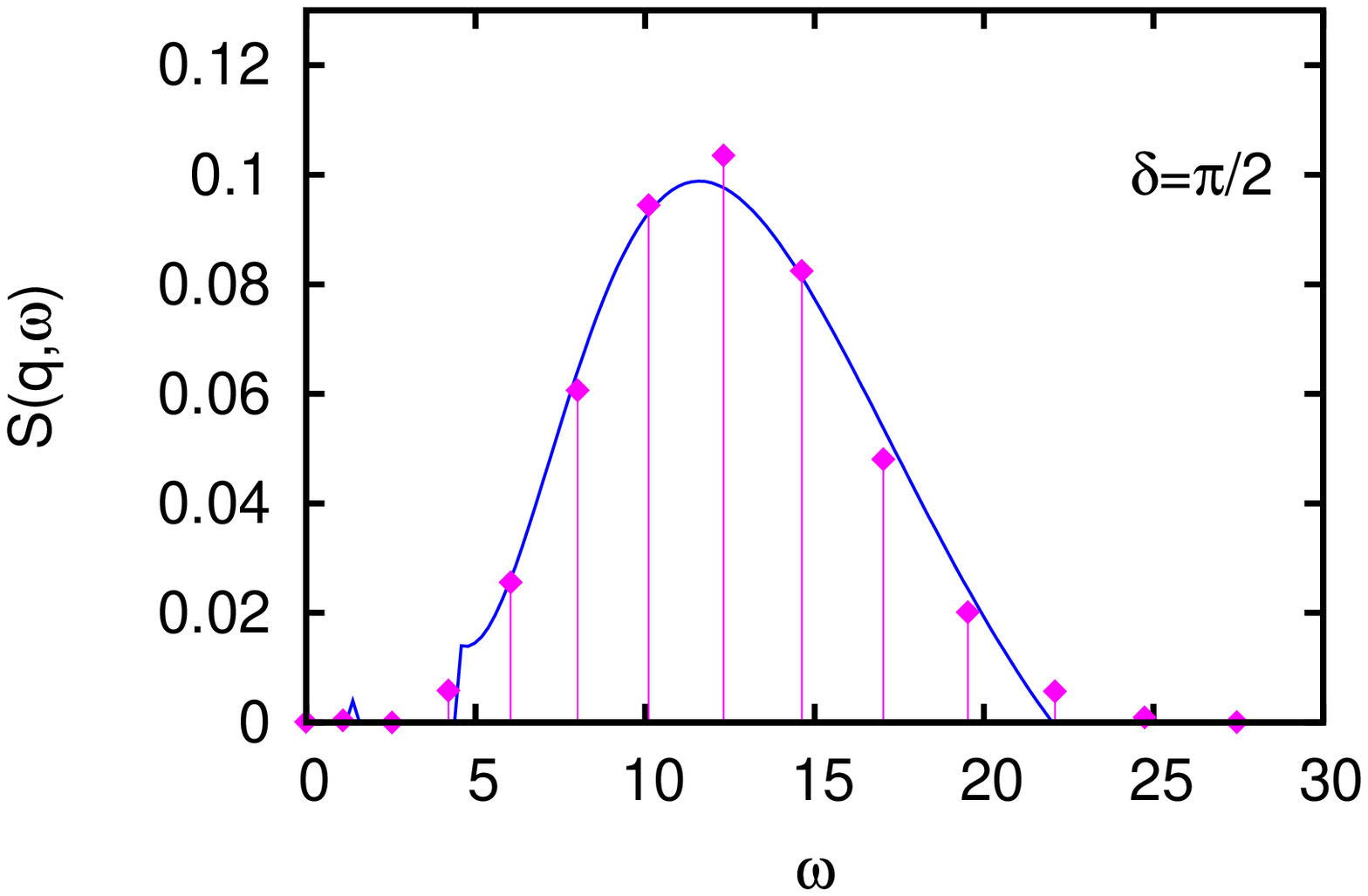}
\includegraphics[width=0.45\textwidth,angle=0]{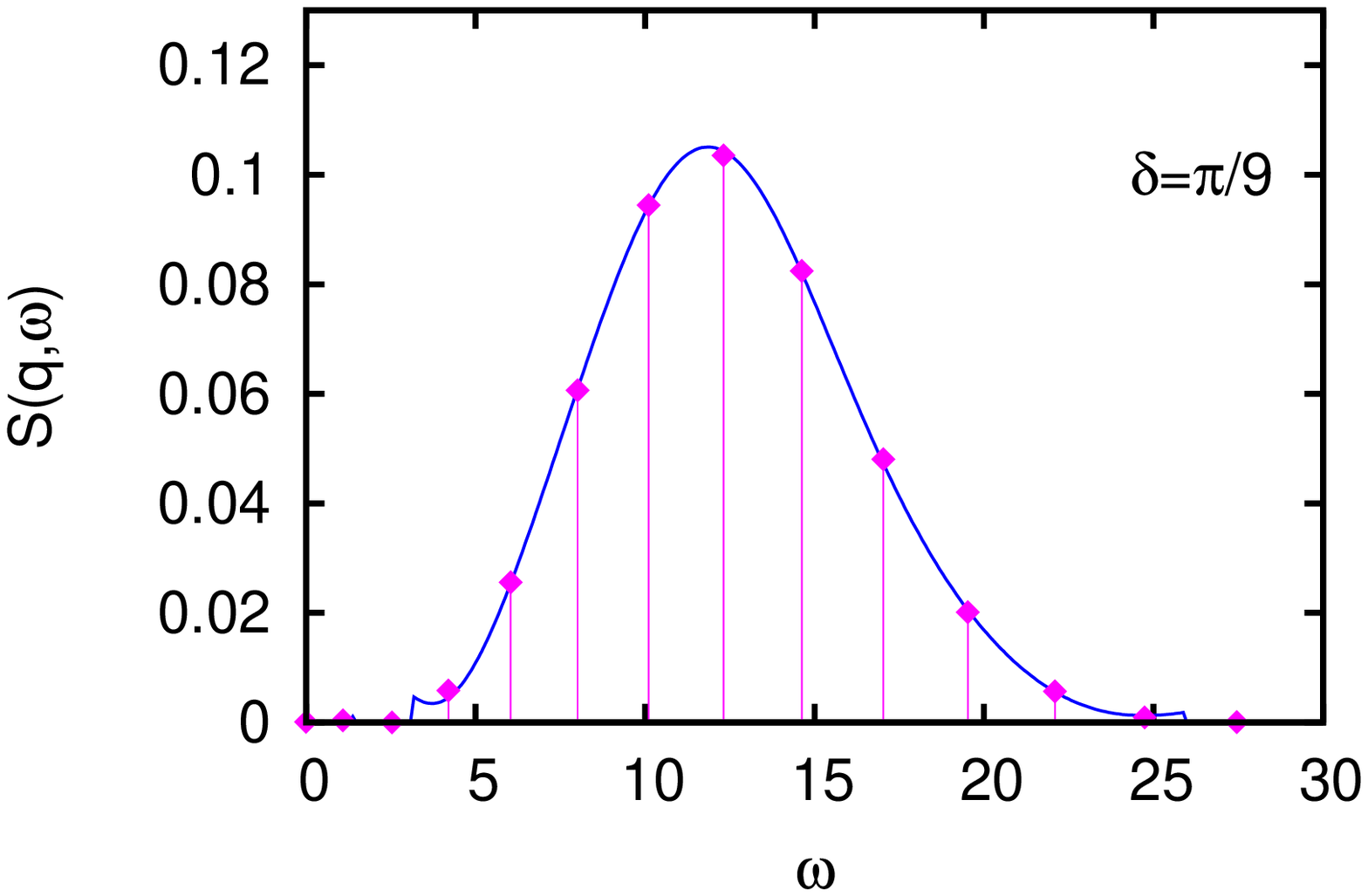}

\caption{(Color online) Dynamic structure function $S(q,\omega)$ for the
AP at $q=5$. Diamonds correspond to the exact values  
and the  curve to
the results derived from the QMC results for $S(q,t_c)$. 
Left panel: imaginary time ($\delta=\pi/2$). Right panel:
complex time ($\delta=\pi/9$).   }
\label{fig:AP_q5}
\end{figure}

When the momentum $q$ increases, the number of modes contributing to
$S(q,\omega)$ also increases, shifting the strength to higher energies. When
$q$ is large enough, the dynamic response is centered around the free atom
recoil energy $\omega_R=\hbar^2 q^2/(2 m)$.~\cite{glyde} We have calculated
the dynamic response for the HP and AP at $q=5$. Our results are reported
in Figs. \ref{fig:HO_q5} and \ref{fig:AP_q5} for HP and AP, respectively.  
The theoretical response shows in both cases, but somehow more clearly in
the HP one, a distribution of modes nearly symmetric around the recoil
energy. 
 The results obtained
for the HP are reported in Fig. \ref{fig:HO_q5} where we compare two cases,
$\delta=\pi/2$ and $\delta=\pi/9$. Our results are shown with a continuous
curve since our
resolution does not allow for a clear separation of the individual excitation
energies. Nevertheless, in the case of using complex time ($\delta=\pi/9$)
the curve precisely reproduces the envelope of the exact spectrum plotted as
vertical lines of strength $h_i$ given by
\begin{equation}
h_i=\frac{1}{\Delta \omega_i}\:\intop_{\omega_i-\Delta
\omega_i/2}^{\omega_i+\Delta \omega_i/2} S(q,\omega)\, d\omega  
\label{exactqw}
\end{equation}
 located at the exact frequency modes $\omega_i$, with $\Delta
 \omega_i=  (\omega_{i+1} -\omega_{i-1})/2$. 
Using just imaginary time produces results which are significantly worse. 
Similar conclusions are drawn from
the results for the AP reported in Fig. \ref{fig:AP_q5}.
The results at $\delta = \pi/2$ are able only to localize the signal of 
$S(q,\omega)$ around $\omega_R$, but they cannot reproduce the shape of the 
spectral function. On the other hand, our results at
$\delta=\pi/9$ match almost perfectly the exact dynamic response.

\section{Conclusions}

The goal of this work is to propose a new QMC strategy aimed at
the study of the dynamic response of quantum systems at zero temperature.
In quantum Monte Carlo methods, the evolution of configurations is
carried out in purely imaginary time, both at zero and  finite
temperatures, in an attempt to describe the main properties of
quantum systems with high accuracy. Unfortunately, dynamics in real time is
not accessible and the usual approach to get information 
on the dynamic response has been to
reconstruct it from purely imaginary-time correlation factors. However, the
ill-posed character of the inverse Laplace transform of noisy data makes
this procedure quite uncertain and with multiple solutions.

Our work is an attempt of progressing in a different way, that is, to
reduce the ill-posed nature of the process by inverting data containing
more information than the smooth signal observed in imaginary-time. Working
in the zero-temperature limit, where quantumness is unavoidable, we have
devised a strategy based on the PIGS method to sample complex-time
correlation factors. Our method consists in the sampling of paths
connecting configurations distributed according to the ground-state wave
function and, in particular, the calculation of the correlation function in complex time
over the sampled paths. The use of high-order actions for the propagation in
complex time has proven to be crucial to get reliable data within a
time window which naturally shrinks when the real axis is approached.
Optimizing the phase $\delta$ of the complex time, we have shown that, in the
two model problems studied, we are able to improve significantly the
calculated dynamic structure factor $S(q,\omega)$. Both at low and high $q$
the description of the dynamics is significantly improved in comparison with the
usual imaginary time approach. Nevertheless, additional effort is needed to confirm 
the usefulness of the 
proposed method to problems in two and three dimensions and with more particles. 
Work is in progress in our group to extend this formalism to many-particle
systems.  

\begin{acknowledgments}
This research was supported under the MICINN-Spain, Grant No.
FIS2011-25275, ERC through the QGBE Grant, and Provincia Autonoma di Trento. 
Additional support was provided by a Grant from the
Qatar National Research Fund  No. NPRP 5-674-1-114.
\end{acknowledgments}

\appendix

\section{}

In this Appendix, we report the explicit expressions for 
the estimator $O_A(x_0,\ldots,x_M)$ appearing in Eq.
\ref{eq:EstimatorPropagator} using different actions and having chosen to 
sample the paths $\{ x_1, x_2, \ldots ,x_{M-1} \}$ with $p_{\text{path}}(x_0, x_1,\ldots,x_M)$ 
defined in Eq. \ref{eq:ppath}. In general, 
$O_A$ is a complex number that can be rewritten in the form
\begin{equation}
O_A(x_0,\ldots,x_M) = \frac{\prod_{k=1}^M G (x_k,x_{k-1};\varepsilon_c)}
{p_{\text{path}}(x_0, x_1,\ldots,x_M)} = \prod_{k=1}^M \frac{G (x_k,x_{k-1};\varepsilon_c)}
{G_{\text{free}}(x_k,x_{k-1};\tau_s)}
\equiv \exp(C) \exp(iA) \ ,
\label{splitcomplex}
\end{equation}
with $\varepsilon_c = \varepsilon_m e^{-i \delta}$. The terms $C$ and 
$A$ are respectively the logarithm of the modulus and the phase of 
the complex number $O_A$, and their formula depends on the approximation 
scheme chosen for the complex-time propagator.

In the primitive action (PA) approximation, introducing the propagator
$G_{\text{PA}}$ in Eq. \ref{splitcomplex} we get 
\begin{equation}\label{eq:Cprim}
C_{\text{PA}} = \sum_{k=1}^M \left[ -\frac{(x_k-x_{k-1})^2}{4 \lambda} 
\left( \frac{\sin \delta }{\varepsilon_m} - \frac{1}{\tau_s} \right) -
\varepsilon_m \frac{V(x_k) 
+ V(x_{k+1})}{2 \hbar} \sin \delta \right]
\end{equation}
and
\begin{equation}\label{eq:Aprim}
A_{\text{PA}} =  \sum_{k=1}^M \left[ \frac{(x_k-x_{k-1})^2}{4 \lambda
\varepsilon_m} \cos \delta  - 
\varepsilon_m \frac{V(x_k) + V(x_{k+1})}{2 \hbar} \cos \delta \right] \ .
\end{equation}

In Chin's approximation (CA) the propagator is given by
\begin{eqnarray}
G_{\text{CA}} & = & \prod_{j=0}^{3} \exp \left( i
\frac{(x_{k,j+1}-x_{k,j})^2}{4 \lambda t_j \varepsilon_c} \right) \exp \left(
-i \frac{V(x_{k,j})+V(x_{k,j+1})}{2 \hbar} v_j \varepsilon_c \right)
\nonumber \\
& & \times \exp \left( i \frac{u_0}{3} \frac{W(x_{k,j}) + W(x_{k,j+1})}{2
\hbar} \varepsilon_c^3 \right) \ ,
\label{chinpropagator} 
\end{eqnarray}
with a generalized potential $W(r)$, due to the double commutator $[\hat{V},[\hat{K},\hat{V}]]$, 
and 
parameters $t_j$, $v_j$, and $u_0$ reported in Ref.
\onlinecite{Sakkos09}.
Introducing this propagator in Eq. \ref{splitcomplex}, we can find the
functions $C_{\text{CA}}$ and $A_{\text{CA}}$,
\begin{eqnarray}
C_{\text{CA}} & = & \sum_{j=1}^4  \left[ \left( -\frac{(x_{k,j+1}-x_{k,j})^2}
{4 \lambda t_j} 
\right) \left( \frac{\sin \delta }{\varepsilon_m} - \frac{1}{\tau_s} \right) + 
\right. \nonumber \\
& & \left( - \varepsilon_m v_j \frac{V(x_{k,j+1}) + V(x_{k,j})}{2 \hbar}
\right) \sin \delta  + 
\nonumber \\
& & \left. \left( \varepsilon_m^3 \frac{u_0}{3} \frac{W(x_{k,j+1}) + W(x_{k,j})}{2 
\hbar} \right) \sin (3 \delta) \right] 
\label{eqca}
\end{eqnarray}
and
\begin{eqnarray}
A_{\text{CA}} & = & \sum_{j=1}^4  \left[ \left( \frac{(x_{k,j+1}-x_{k,j})^2}
{4 \lambda t_j \varepsilon_m} \right) \cos \delta + \right. \nonumber \\
& & \left( - \varepsilon_m v_j \frac{V(x_{k,j+1}) + V(x_{k,j})}{2 \hbar} \right) 
\cos \delta  + \nonumber \\
& & \left. \left( \varepsilon_m^3 \frac{u_0}{3} \frac{W(x_{k,j+1}) + W(x_{k,j})}
{2 \hbar} \right) \cos(3 \delta) \right] 
\end{eqnarray}
Unfortunately, for $\delta<\pi/3$, the term with
$\varepsilon_m^3$ in the expression of $C_{\text{CA}}$  (\ref{eqca}) is
positive, and then $\exp(C_{\text{CA}})$ can become exceedingly large and
spoil the calculation.

In order to circumvent this problem, we have worked with the sixth-order
expansion~\cite{Zillich10} 
\begin{eqnarray}
e^{\varepsilon_c \hat{H}} & \simeq & 
\frac{64}{45} e^{\varepsilon_c\hat{V}/8} e^{\varepsilon_c\hat{K}/4} 
e^{\varepsilon_c \hat{V}/4}e^{\varepsilon_c \hat{K}/4}e^{\varepsilon_c \hat{V}/4} 
e^{\varepsilon_c \hat{K}/4}
e^{\varepsilon_c \hat{V}/4}e^{\varepsilon_c \hat{K}/4}e^{\varepsilon_c \hat{V}/8} \nonumber \\
& & - \frac{4}{9} e^{\varepsilon_c \hat{V}/4} e^{\varepsilon_c \hat{K}/2} 
e^{\varepsilon_c \hat{V}/2}
e^{\varepsilon_c \hat{K}/2}e^{\varepsilon_c \hat{V}/4} + \frac{1}{45} 
e^{\varepsilon_c \hat{V}/2}
e^{\varepsilon_c \hat{K}}e^{\varepsilon_c \hat{V}/2} \ , 
\label{eq:zillich}
\end{eqnarray}
which is built without double-commutator terms. This expansion 
corresponds to a linear combination of expansions approximated 
with PA over the same time $\varepsilon_c$ but with different time 
steps (precisely, $\varepsilon_c/4$ in the first term, $\varepsilon_c/2$ 
in the second term and $\varepsilon_c$ in the third term). Therefore, 
the complete formula for the exponent $C_{\text{ZA}}$ and for the 
phase $A_{\text{ZA}}$ in the Zillich approximation are easily obtained 
from $C_{\text{PA}}$ and $A_{\text{PA}}$ (Eqs. \ref{eq:Cprim} 
and \ref{eq:Aprim}) calculated for different values of $\varepsilon_c$.

\section{}
\begin{figure}

\includegraphics[width=0.85\textwidth,angle=0]{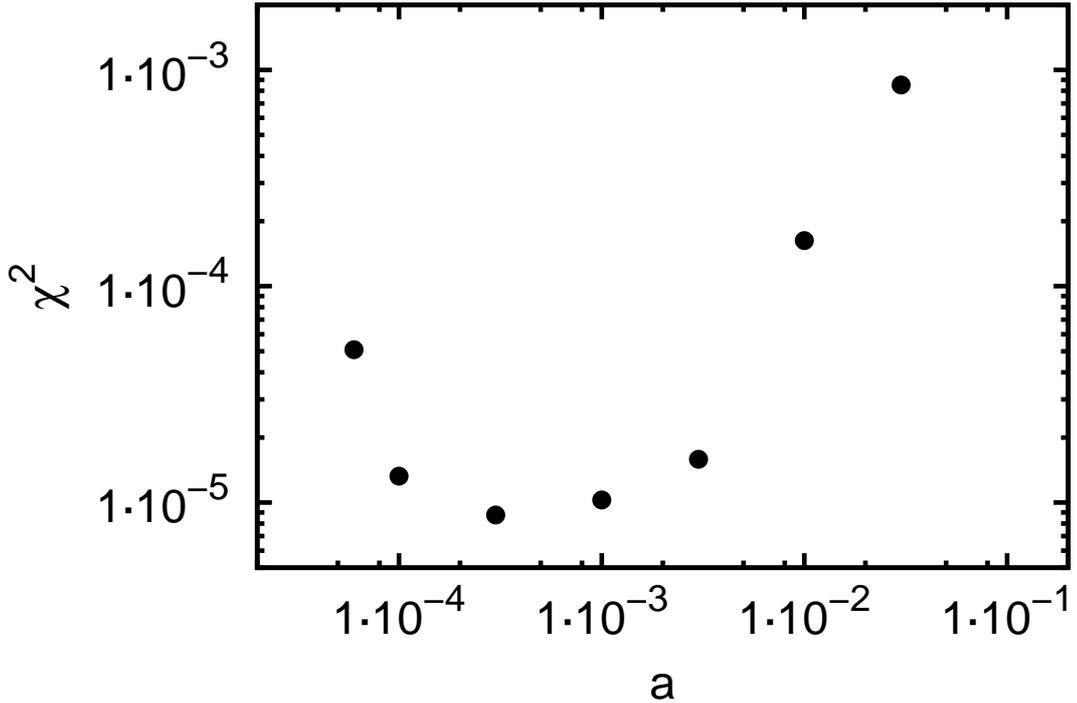}

\caption{Residual $\chi^{2}$ between $C_{\text{QMC}}(t_c)$ and
$C_{\text{INV}}(t_c,a)$ as a function of the
regularization parameter $a$. The data corresponds to  the 
calculation of the density correlation function $S(q,t_c)$ (Eq. \ref{sqtc}) 
in complex time $t_c = t_m e^{-i\delta}$ for the AP 
at $q=1.5$ and $\delta=\pi/4$.} 
\label{fig:regularitzador}
\end{figure}

We discuss in this Appendix the method that we have followed to find the
optimal regularization parameter (see Sec. II.B). 
 Given the spectral
function $S_{\text{INV}}(\omega,a)$ obtained inverting a series of QMC data for
the complex-time correlation function $C_{\text{QMC}}(t_c)$ with a certain
regularization parameter $a$, we calculate the complex-time correlation
function $C_{\text{INV}}(t_c,a)$ obtained from the integral transform of
$S_{\text{INV}}(\omega,a)$:
\begin{equation}
C_{\text{INV}}(t_c,a) = \int d\omega e^{-i t_c \omega} S_{\text{INV}}(\omega,a) \ . 
\end{equation}
Then we calculate the residual $\chi^{2}$ between $C_{\text{QMC}}(t_c)$ and
$C_{\text{INV}}(t_c,a)$ as a function of the regularization parameter $a$. 
When $a$ is large, the regularization procedure modifies the inversion process 
up to the point that $C_{\text{INV}}(t_c,a)$ starts to differ from the previous 
Monte Carlo data $C_{\text{QMC}}(t_c)$, thus showing an increase in
$\chi^{2}$. 
For very
small $a$, the noise in the Monte Carlo data is largely amplified and
the inversion procedure itself starts to produce meaningless results,
giving rise once again to the increase in $\chi^{2}$.
A plot of the total residual $\chi^{2}$ versus the regularization parameter $a$ shows a 
minimum,
as shown in Fig. \ref{fig:regularitzador} 
for the case of the AP  data at $q=1.5$ and $\delta=\pi/4$.

In the best scenario, with high quality Monte Carlo data, an optimal
regularization parameter may allow avoiding both problems. In any
case, the full inspection of the inversion landscape for several values of
the regularization parameter is a quick calculation.

\end{document}